\documentclass[aps,10pt,notitlepage,superscriptaddress,pra,twocolumn]{revtex4-1}

\usepackage{amsmath}
\usepackage{amsfonts}
\usepackage{amssymb}
\usepackage{graphicx}
\usepackage{sidecap}
\usepackage{braket}

\usepackage{blindtext}
\usepackage{lipsum}
\newcommand*\diff{\mathop{}\!\mathrm{d}}

\usepackage{enumitem}
\usepackage{appendix}
\usepackage{subfigure}
\usepackage{color}
\usepackage[dvipsnames]{xcolor}
\usepackage{physics}
\usepackage{sidecap}

\usepackage[ruled,vlined,noline]{algorithm2e}

\usepackage{natbib}
\usepackage{amsthm}
\theoremstyle{plain}

\theoremstyle{definition}

\theoremstyle{remark}

\usepackage{enumerate}

\usepackage{comment}

\usepackage{hyperref}

\usepackage[capitalize,noabbrev]{cleveref}

\begin{document}

\title{Understanding and controlling the geometry of memory organization in RNNs}

\author{Udith Haputhanthri}
\thanks{Co-first authors}
\affiliation{CNC Program, Stanford University, Stanford}
\author{Liam Storan}
\thanks{Co-first authors}
\affiliation{CNC Program, Stanford University, Stanford}
\author{Yiqi Jiang}
\thanks{Co-first authors}
\affiliation{CNC Program, Stanford University, Stanford}
\affiliation{James H. Clark Center for Biomedical Engineering \& Sciences, Stanford University, Stanford}
\author{\\ Tarun Raheja}
\affiliation{Kipo AI, San Francisco}
\author{Adam Shai}
\affiliation{Simplex, Astera Institute, Emeryville, CA}
\author{Orhun Akengin}
\affiliation{CNC Program, Stanford University, Stanford}
\author{Nina Miolane}
\affiliation{Geometric Intelligence Lab, UC Santa Barbara, Santa Barbara}
\author{\\ Mark J. Schnitzer}
\thanks{Supervisors}
\affiliation{CNC Program, Stanford University, Stanford}
\affiliation{James H. Clark Center for Biomedical Engineering \& Sciences, Stanford University, Stanford}
\affiliation{Howard Hughes Medical Institute}
\author{Fatih Dinc}
\thanks{Supervisors}
\affiliation{CNC Program, Stanford University, Stanford}
\affiliation{Physics \& Informatics Laboratories, NTT Research Inc., Sunnyvale, CA 94085, USA}
\affiliation{Geometric Intelligence Lab, UC Santa Barbara, Santa Barbara}
\affiliation{Kavli Institute for Theoretical Physics, UC Santa Barbara, Santa Barbara}

\author{Hidenori Tanaka}
\thanks{Supervisors}
\affiliation{Physics \& Informatics Laboratories, NTT Research Inc., Sunnyvale, CA 94085, USA}
\affiliation{CBS-NTT Program in Physics of Intelligence, Harvard University, Cambridge, MA 94305, USA}

\begin{abstract}
Training recurrent neural networks (RNNs) is a high-dimensional process that requires updating numerous parameters. Therefore, it is often difficult to pinpoint the underlying learning mechanisms. To address this challenge, we propose to gain mechanistic insights into the phenomenon of \emph{abrupt learning} by studying RNNs trained to perform diverse short-term memory tasks. In these tasks, RNN training begins with an initial search phase. Following a long period of plateau in accuracy, the values of the loss function suddenly drop, indicating abrupt learning. Analyzing the neural computation performed by these RNNs reveals geometric restructuring (GR) in their phase spaces prior to the drop. To promote these GR events, we introduce a temporal consistency regularization that accelerates (bioplausible) training, facilitates attractor formation, and enables efficient learning in strongly connected networks. Our findings offer testable predictions for neuroscientists and emphasize the need for goal-agnostic secondary mechanisms to facilitate learning in biological and artificial networks.
\end{abstract}

\maketitle

\section{Introduction}

Recurrent neural networks (RNNs) play a crucial role in studies of biological and artificial neural systems \cite{Durstewitz2022, perich2021inferring,sussillo2015neural,valente2022extracting,finkelstein2021attractor,langdon2023unifying}. As universal approximators \cite{schafer2006recurrent,beiran2021shaping}, RNNs can be trained to perform complex associative computations and form memory-subserving subspaces \cite{mante2013context,Vlachas2020-ld,masse2019circuit}. Fortunately, unlike many contemporary artificial intelligence algorithms \cite{adadi2018peeking}, the computations performed by RNNs can often be reverse-engineered using tools from dynamical systems theory \cite{maheswaranathan2019reverse,sussillo2013opening,Golub2018}. This key aspect of interpretability has enabled a large body of neuroscience work to utilize RNNs as trainable dynamical systems and to study the emergent neural algorithms employed to solve many neuroinspired tasks \cite{perich2021inferring,valente2022extracting,finkelstein2021attractor,perich2020rethinking,Khona2022, Levenstein1074, NEURIPS2023_50d6dbc8,Yoon2013,Chaudhuri2019,redish1996,rajan2016recurrent,duncker2021dynamics,cohen2020recurrent}. However, given the high-dimensional and complex nature of learning in RNNs \cite{NEURIPS2022_495e55f3,pascanu2013difficulty,doya1992bifurcations}, probing the emergence of these neural algorithms during training is often a challenging task.

A recent promising direction has focused on modeling nonlinear dynamical systems using interpretable and analytically tractable RNN architectures \cite{mante2013context, schmidt2021identifying, koppe2019identifying}. This approach primarily relies on the assumption that many real-world tasks can be performed with relatively simple attractor landscapes, \textit{i.e.}, stable states in RNNs' phase spaces that nearby trajectories converge to \cite{maheswaranathan2019reverse,sussillo2013opening,dubreuil2022role,yang2019task,eisenmann2023bifurcations}. However, one of the primary challenges in training RNNs lies in the fact that weight initialization often fails to place the network in a weight subspace with the desired attractor landscapes. Consequently, the learning process must navigate through subspaces with distinct (often complicated) geometrical properties. Thus, even if the final network converges to a simple solution, extracting learning mechanisms requires studying RNNs' complicated phase spaces in a task- and architecture-agnostic manner.

A key insight to address this challenge may lie in the changing phase space geometries during spikes or jumps in loss curves. These special events signal substantial changes in network dynamics and are frequently observed during task training \cite{eisenmann2023bifurcations,delmastro2023dynamics}. To date, these abrupt changes have often been studied in conjunction with bifurcations \cite{Monfared2020,eisenmann2023bifurcations,HASCHKE200525,delmastro2023dynamics,hess2023generalized}, events that mark qualitative shifts in system behavior, such as changes in fixed (equilibrium) points or periodic orbits \cite{strogatz2018nonlinear}. Yet, as we will see, such abrupt geometric restructuring in the phase space can occur without bifurcations. This observation, therefore, reinforces the need for an agnostic study of changes in phase space geometries during learning. Moreover, to date, the harmful effects of these abrupt transitions, such as exploding gradients, have received considerable attention \cite{pascanu2013difficulty,NIPS2017_f2fc9902,rehmer2022, pmlr-v108-ribeiro20a}. However, their constructive role in guiding high-dimensional RNNs to train effectively in a general setting remains largely unexplored.

In this work, we study diverse network architectures and short-term memory tasks and conduct a thorough scientific investigation of the sudden learning phenomena observed in RNNs. We observe two distinct phases during the training of RNNs, search and comprehension, divided by a sudden jump in the loss function values. Examining these transitions reveals the crucial and constructive role of geometric restructuring in phase space as precursors to abrupt learning. We refer to these emergent phase space structures with slow time dynamics as ``computational structures,'' which presumably subserve computation in short-term memory tasks. Inspired by these observations, we next propose the temporal consistency regularization (TCR), a spatio-temporally local mechanism that incentivizes the formation of such computational structures. We find that TCR accelerates training by shortening the search phase, and can facilitate online learning of fixed point attractors. Surprisingly, TCR enables training networks within a strongly connected regime, which are known to become chaotic in the limit of infinite neurons \cite{sompolinsky1988chaos}. Such a feat was considered rather difficult with the backpropagation through time algorithm and that had led to the birth of new paradigms such as reservoir computing \cite{maass2002real,jaeger2004harnessing}, FORCE \cite{sussillo2009generating} and full-FORCE \cite{depasquale2018full}. 

Our work suggests a nuanced interpretation of abrupt learning with three key insights: i) geometric restructuring during learning (often studied in the form of bifurcations, though see our recent work \cite{dinc2025ghost}) is not always a simple inconvenience that needs to be mitigated, ii) distinct mechanisms other than bifurcations can facilitate the growth of computational structures and subsequent abrupt learning, and iii) enforcing temporal consistency enables faster training in diverse RNN architectures and short-term memory tasks. Overall, our work constitutes a rigorous systematic study of learning dynamics in high-dimensional dynamical systems, links formation of (complex) computational structures to abrupt drops in loss function values, and our findings have direct implications for learning in biological and artificial neural systems.

\section{Results}
\label{sec:results}

\begin{figure*}
    \centering
    \includegraphics[width=0.9\textwidth]{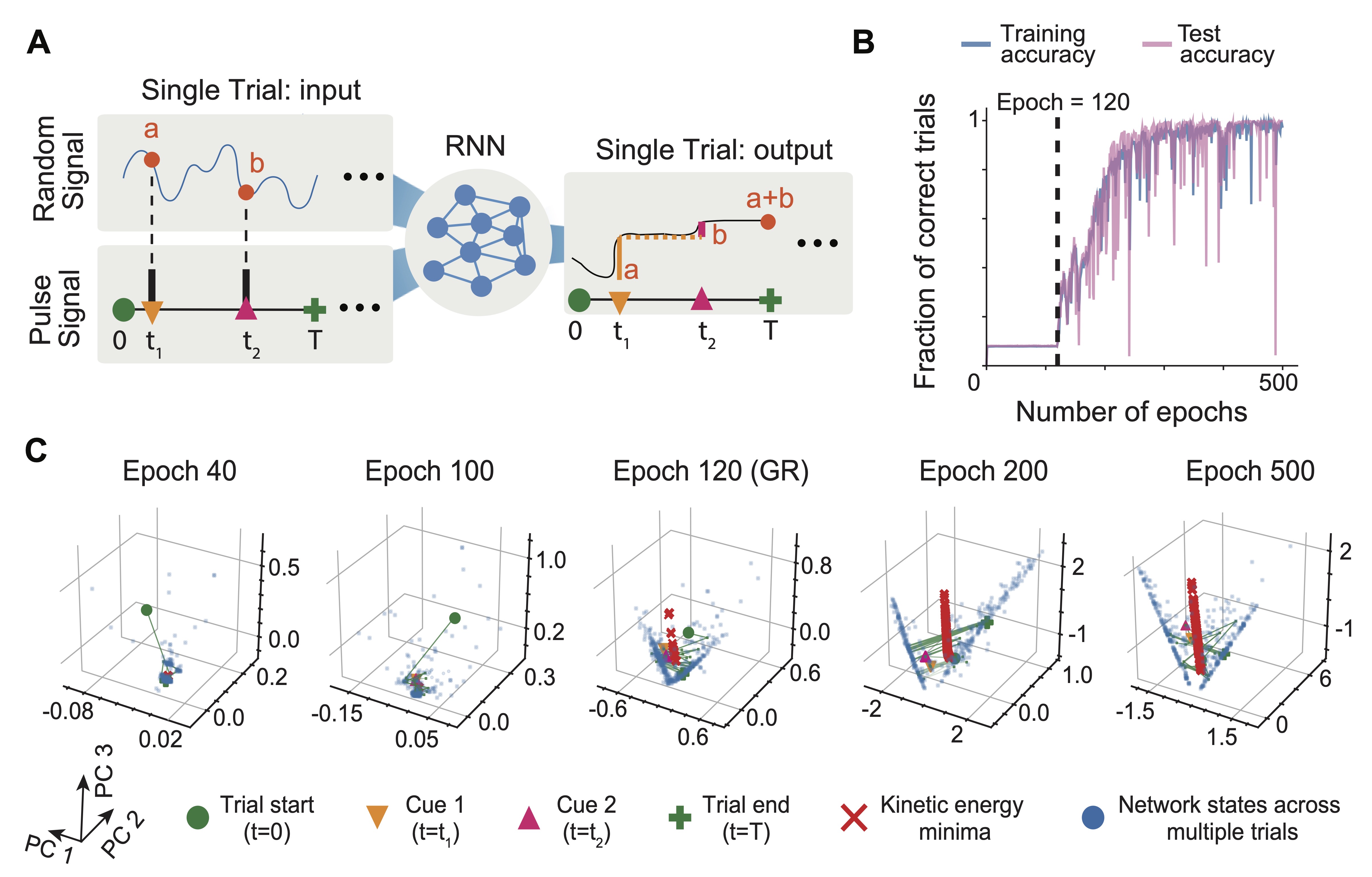}
    \caption{\textbf{Geometric restructuring (GR) subserves the sudden accuracy jump during the learning of the delayed addition task.}
\textbf{A} In the delayed addition task, the network receives signals from two channels: the first channel has continuously valued random signals, the second channel is zero for almost all times, but has two binary pulses at times $t_1$ and $t_2$. During these pulse times, the network needs to memorize the ``cue'' values presented in the first channel and must output their sum at the end of the trial ($t=T$). \textbf{B} The fraction of correct trials as a function of number of training epochs. The training and testing curves both show two phases, with a sudden transition at the epoch 120. \textbf{C} Top three principal components of the network's activations over multiple trials (blue), neural trajectories during a single trial (green), and kinetic energy minima corresponding to slow points (red). A geometric restructuring of slow points at epoch 120 separates the search and comprehension phases of learning. For \textbf{B-C}, we used a representative unregularized network with $N=40$ neurons trained to perform a delayed addition task ($500$ epochs, $T=40$).}
    \label{fig:fig1}
\end{figure*}

\subsection{Abrupt learning in recurrent neural networks during task-training}  \label{sec:sec2p1}

To illustrate the abrupt learning phenomenon, we start our analysis with piecewise linear recurrent neural networks (PLRNNs). The PLRNNs follow the discretized time evolution equations \cite{schmidt2021identifying}:
\begin{equation}
\label{eq:ds}
    x[t]= Ax[t-1] + W\phi(x[t-1]) + Cs[t] + h, 
\end{equation}
where $x[t] \in \mathbb{R}^{N}$ are the state variables, $r[t] = \phi(x[t])$ the firing rates of $N$ neurons with the $\phi(.)$ non-linearity, $A \in \mathbb{R}^{N \times N}$ a diagonal matrix setting the decay times of neural activities, $W \in \mathbb{R}^{N \times N}$ the recurrent connectivity matrix, $s[t] \in \mathbb{R}^{N_{\rm in}}$ is the external input, $C \in \mathbb{R}^{N \times N_{\rm in}}$ input weights, and $h \in \mathbb{R}^{N}$ the biases. Here, piecewise linear specifically refers to the RELU nonlinearity, \textit{i.e.}, $\phi(z) = \text{max}(0, z)$. The network output is defined as $\hat o(t) = W_{\rm out} x(t) + h_{\rm out}$ for $W_{\rm out}\in \mathbb R^{N_{\rm out} \times N}$ and $h_{\rm out} \in \mathbb R^{N_{\rm out}}$. Unless otherwise specified, all parameters ($A$, $W$, $C$, $h$, $W_{\rm out}$, $h_{\rm out}$) are learned during training where PLRNNs aim to perform predefined tasks. 

In this work, we will eventually consider a wide range of tasks. Here, as a simple and illustrative example, we start by focusing on delayed addition. In this short-term memory task, there are two channels that provide input to the network (Fig. \ref{fig:fig1}\textbf{A}). Within each trial of length $T$, the first channel ($u_1(t)$) contains continuously valued random ``cues'', while the second channel is mainly zero ($u_2(t)=0$), except for two pulses at two distinct times ($u_2(t_1)=u_2(t_2)=1$). The output of the task corresponds to the sum of the two cues at times $t_1$ and $t_2$, \textit{i.e.}, $o(T)=u_1(t_1)+u_1(t_2)$, which should be returned by the trained network at the end of the trial (Fig. \ref{fig:fig1}\textbf{A}). Notably, in order to perform this task accurately, networks should be able to hold and manipulate numbers in short-term memory for $O(T)$ durations.

When we train PLRNNs to perform this task, we observe two phases of learning, namely the search and comprehension phases (Fig. \ref{fig:fig1}\textbf{B}). During the search phase, the network does not show a significant performance improvement in either training or test trials. However, at epoch 120, the representative network enters the comprehension phase, in which the training performance abruptly improves. Our goal for the rest of this work is to understand and facilitate the mechanisms behind such abrupt (later revealed to be geometrical) transitions that signal skill acquisition (\textit{e.g.}, here, acquiring the capability to hold and manipulate short-term memory content).

\subsection{Geometric restructuring underlies skill acquisition during learning}
To understand the mechanisms behind abrupt changes during network training, we draw upon dynamical systems theory. This theory has been essential in analyzing RNNs, where researchers focus on identifying attractor structures \cite{eisenmann2023bifurcations,schmidt2021identifying,delmastro2023dynamics} - which can take various forms such as lines \cite{mountoufaris2024line}, rings \cite{kim2017ring}, or toroids \cite{gardner2022toroidal}. The parameter space containing these attractors is partitioned into distinct subspaces, each with unique geometrical properties. Transitions between these subspaces involve fundamental changes in the geometric landscape, which in turn determine the computational capabilities of the network. In this work, we refer to these transitions as geometric restructuring events, though they often manifest as and were studied in conjunction with bifurcations \cite{eisenmann2023bifurcations,pmlr-v202-hess23a,strogatz2018nonlinear}.

In a general setting, novel geometrical properties may emerge during training without explicit bifurcations. For instance, our recent work has shown that RNNs can learn to delay their actions by developing slow points \cite{dinc2025ghost}, states for which $\dot x[t] \neq 0$ but still are capable of drawing nearby trajectories towards themselves. Consequently, in this work, we adopt a bifurcation-agnostic methodology to study changes in geometrical landscapes learned by RNNs during training. One such method, used in a previous seminal work \cite{ribeiro2020beyond}, involves extracting ``proxy bifurcation diagrams'' by letting the trained networks produce trajectories for longer times and noting the eventual states that the networks arrive to, \textit{e.g.}, fixed point attractors. This approach is limited by its inability to distinguish between slow points and fixed points. This distinction is crucial, as failing to make it can lead us to miss or misidentify the fundamental mechanisms that give rise to new computational capabilities - the central focus of our study. 

An alternative approach to identifying attractor structures in the phase space involves minimizing a function of state changes, also known as the ``kinetic energy of the network'' \cite{sussillo2013opening}. Such an approach is also well poised to identify non-attractive structures in the phase space, \textit{e.g.}, saddle nodes that have $\dot x[t] = 0$ but do not necessarily attract nearby states. However, this method is numerical in nature and requires additional analytical assistance to fully distinguish slow points from fixed points. The third and final approach involves studying the evolution of so-called latent circuits during training \cite{dinc2025ghost}, which is an analytical method but applies to a limited class of RNN architectures. In this work, thanks to its agnosticism to network architectures and the type of attractor structures in the phase space, we mostly use the second approach. Yet, we supplement it with the third approach to study a representative example, which uncovers that the identified geometrical structures are not necessarily fixed points but can be slow points. Therefore, changes in their numbers or properties should not be directly referred to as bifurcations.

We operationalize the identification of geometrical landscapes supported by a particular weight matrix, $W$, with an energy minimization approach \cite{sussillo2013opening,golub2018fixedpointfinder}: 
\begin{equation} \label{eq:slow_points}
    x^* \in \arg \min_{x \in \mathcal N(x_0)} E(x) = \arg \min_{x \in \mathcal N(x_0)} \left|\left|\frac{\partial x[t]}{\partial t}\right|\right|_2^2,
\end{equation}
in which the ``kinetic energy'' of neural trajectories is minimized in the neighborhoods, $\{\mathcal N(x_0)\}$, of few pre-defined state variables, $\{x_0$\}, to identify states with locally minimal speeds ($x^*$, Figure \ref{fig:figs1}\textbf{A-B}), \textit{i.e.}, slow points. Such states with exactly zero speed are referred to as fixed points. In this work, we sample $x_0$ from transient neural activities and minimize locally using a Pytorch-based model (\cite{Paszke2017AutomaticDI}, \textbf{Methods}).

Using this approach, we identified the slow points of the PLRNNs and visualized their time evolution as training progressed (Fig. \ref{fig:fig1}\textbf{C}). As expected, we observed a sudden change in neural activities and the overall slow point structure around epoch 120 (Fig. \ref{fig:fig1}\textbf{C}), signaling a change in the geometrical landscape. This shift was then followed by an abrupt increase in both training and testing accuracies. Notably, the restructuring of the slow point landscape (the geometrical restructuring, GR) occurred in only a few epochs, and the emergent structure was fine-tuned in the rest of the epochs to meet the task’s requirements (Fig. \ref{fig:fig1}\textbf{C}, epochs 120 vs 500). This suggests that the restructuring at epoch 120 propelled the network into a weight subspace that contained the ``right'' geometrical properties. Though this observation alone cannot explain why the network performance suddenly started improving, it is an important step forward as we discuss in Section \ref{sec:two_phases}.

\begin{figure*}[!t]
    \centering
    \includegraphics[width = \textwidth]{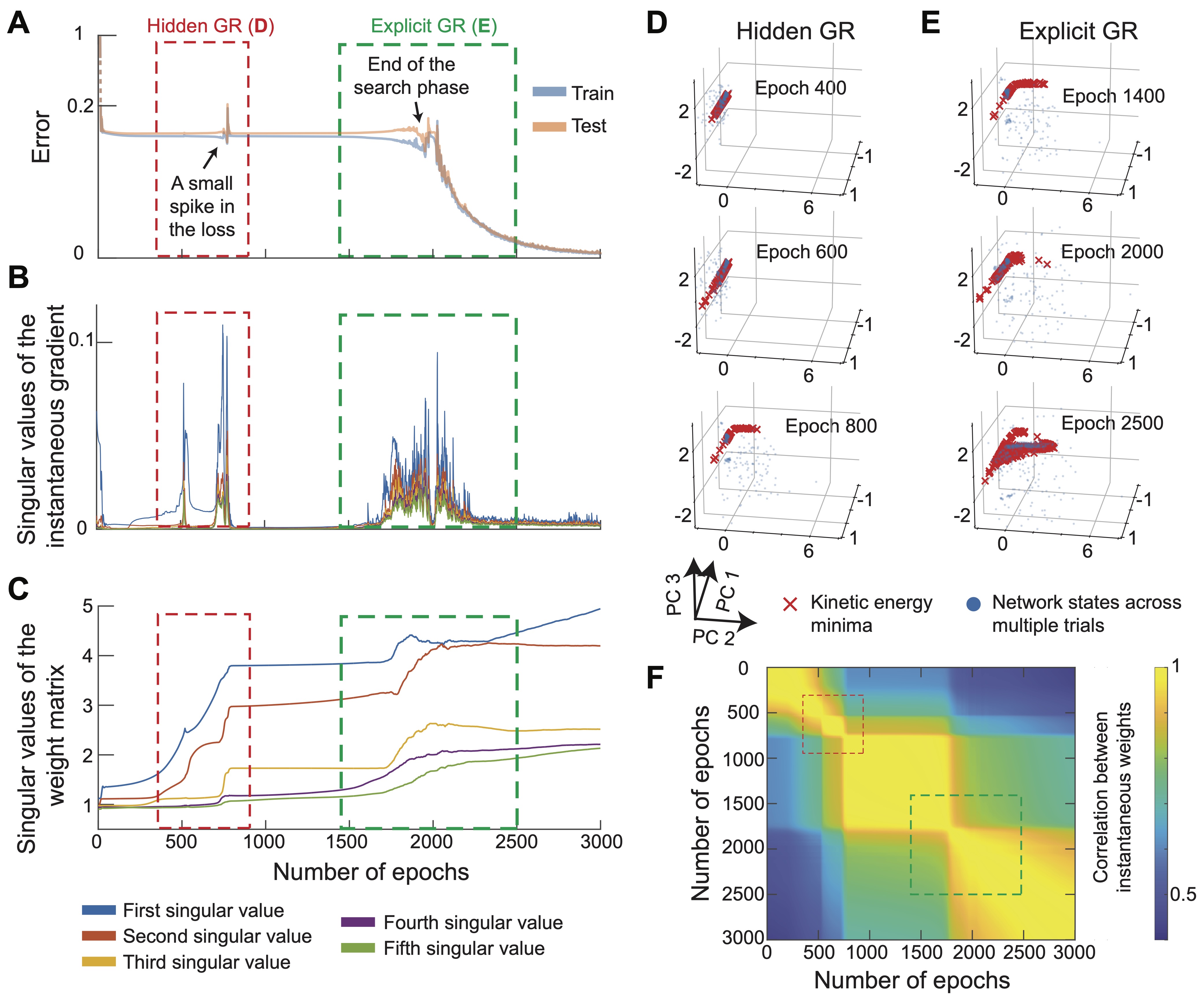}
    \caption{
    \textbf{Geometric restructuring occurs abruptly and forms attractive subspaces during learning.} \textbf{A} Spikes or jumps in the loss function may indicate geometric restructuring. \textbf{B, C} Instabilities in the singular values of the gradient (\textbf{B}) and the weight matrix (\textbf{C}) coincide with abrupt changes in the phase space. \textbf{D, E} Slow point structures emerge in the phase space after hidden (\textbf{D}, Epoch 800) and explicit (\textbf{E}, Epoch 2500) geometric restructuring. These structures are not always attractive (Fig.~\ref{fig:figs1}\textbf{(A-B)}). \textbf{F} Correlation between weights at two distinct time points during training shows high correlations for sustained time intervals, though the weight matrix abruptly changes during GR events. For all panels, we used a representative unregularized network with $N=40$ neurons trained to perform the delayed addition task ($3000$ epochs, $T=40$).}
    \label{fig:fig2}
\end{figure*}

\subsection{Sudden geometrical restructuring is accompanied by destabilization of learning}

In the previous section, we considered a network that was already initialized close to the desired geometrical subspace, evident by the relatively brief search phase (120 epochs). The emergence of the new slow point landscape was accompanied by a jump in loss function and had lasting effects on training performance, resulting in a sustained loss decrease. We categorize such events as ``explicit GR.'' However, not all substantial shifts in the state space geometry may lead to sustained decreases in the loss function. These changes may instead manifest subtly as brief spikes in the loss function, after which the loss function values quickly return to baseline levels.

To illustrate such hidden changes, we considered a training instance, in which the network spent around $2500$ epochs in the search phase and there was a small spike in the learning curve that did not improve or hurt the network performance (Fig. \ref{fig:fig2}\textbf{A}, red rectangle). Yet, it was not clear whether this was a self-correcting behavior or a beneficial event towards the final solution. Therefore, as a first test, we computed the singular value decomposition of the instantaneous gradient and the current weight matrix, $W(t)$, at the training epoch $t$. (As a side note, throughout this work, we use the word ``instantaneous gradient'' to refer to the quantity $W(t)-W(t-1)$, which may be different from the exact gradient as we use the ADAM optimizer \cite{adamOpt_kingma2017adam}.) As the loss function spiked, the gradient became unstable, exhibiting sudden changes (Fig. \ref{fig:fig2}\textbf{B}, red rectangle), and the weight matrix had abrupt, but continuous and sustained, changes (Fig. \ref{fig:fig2}\textbf{C}, red rectangle); ruling out a self-recovery explanation. Specifically, the network did not enter a destabilization region and promptly came back to the prior geometric subspace, instead the learning took the network parameters through different subspaces with distinct geometrical properties \emph{and} learning destabilized during the transition, \textit{i.e.}, a GR event \emph{hidden} in the small spike of the loss function.

Next, we studied the qualitative changes in the slow point landscape (Fig. \ref{fig:fig2}\textbf{D}, Epoch 800). Even though this hidden shift did not immediately improve performance, it served as a step towards finding the desired state-space geometry (see green rectangles in Figs. \ref{fig:fig2}\textbf{A-C}, and Fig. \ref{fig:fig2}\textbf{E}). In this example, the GR at epoch 800 marked the emergence of a second ``arm'' of slow points (Fig. \ref{fig:fig2}\textbf{D}) that subsequently became a plane-like slow point structure (Fig. \ref{fig:fig2}\textbf{E}). The common elements of both GR events were the sudden destabilization of the gradients and the continuous changes in the weight matrices (Figs. \ref{fig:fig2}\textbf{B, C}), but the former was hidden in the sense that no immediate sustained changes were visible in the training curve (Fig. \ref{fig:fig2}\textbf{A}). Finally, when we visualized the changes in the weight matrix throughout the training (Fig. \ref{fig:fig2}\textbf{F}), we observed a block diagonal structure. The network experienced minimal changes in weight values between, but abrupt adjustments during, GR events. After the second GR event, the weights were gradually (but still abruptly) updated, accompanied by a jump in the loss values.

So far, we have demonstrated on two illustrative examples (Figs. \ref{fig:fig1} and \ref{fig:fig2}) that passing through the GR events may endow networks with novel computational abilities. These events were enabled by structural changes in the geometrical landscapes of the networks, signaled by large fluctuations in the gradients and abrupt and sustained changes in the weight matrices (Figs. \ref{fig:fig2}\textbf{B, C}). Examining several networks trained under identical conditions revealed that both phenomena were consistently present alongside the jumps in the loss function (Fig. \ref{fig:figs1}\textbf{C-D}). The singular values of both quantities showed statistically significant differences when comparing the GR event period (averaged over $\pm50$ epochs) to the post-event period ($400\pm 50$ epochs after; Wilcoxon signed rank test, n=19 networks, $p<10^{-3}$). Finally, studying these jumps in these networks, we visually confirmed GR events (data not shown), generalizing our analyses beyond two examples.

\subsection{The emergence of slow points underlies abrupt skill acquisition}
We now study the mechanisms underlying abrupt skill acquisition using interpretable methods that utilize a particular class of (rank-one) RNNs:
\begin{equation}
    \tau \dot r(t) = -r(t) + \tanh( \frac{1}{N}mn^T r(t) + C u(t) + h),
\end{equation}
where $r(t) \in \mathbb R^N$ refer to firing rates and the recurrent weight matrix, $W = \frac{1}{N} mn^T$, is constrained to be rank one with a re-parametrization via $m \in \mathbb R^N$ and $n \in \mathbb R^N$. In these architectures, one can define a one-dimensional latent variable, $\kappa(t) = \frac{n^T r(t)}{N}$, which follows a closed dynamical system equation:
\begin{equation}
    \tau \dot \kappa(t) = -\kappa(t) + \frac{n^T}{N}\tanh( m\kappa(t) + C u(t) + h).
\end{equation}
We refer to this dynamical system with the state variable $\kappa(t)$ as the ``latent circuit,'' whose evolution during learning can be analytically studied. For example, the slow points in this structure are straightforward to obtain by finding $\kappa$ for which $\dot \kappa \approx 0$. More detailed information is also available since $\dot \kappa(t)$ has a clear analytical form and can be computed once the network parameters are known.

\begin{figure*}
    \centering
    \includegraphics[width=0.7\textwidth]{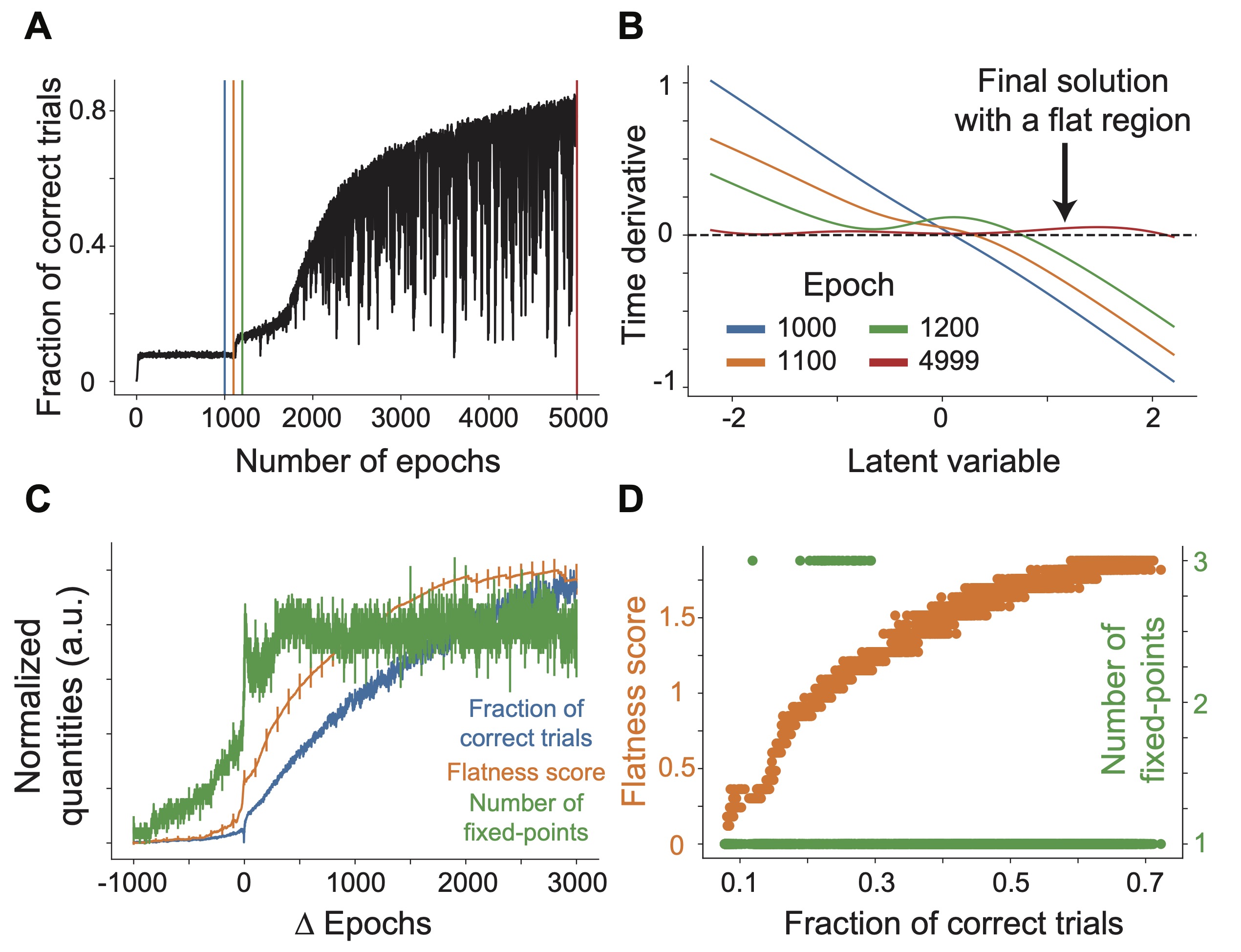}
    \caption{\textbf{The emergence of new geometric structures are not necessarily preceeded by bifurcations.} To study the emergence of computational structures in the delayed addition tasks, we studied rank-one RNNs with interpretable latent circuits. In these RNNs, the recurrent weight matrix is enforced to have only one non-zero singular value, which allows defining a one-dimensional latent circuit with its own autonomous dynamical equations that underlie the computation. We trained unregularized rank-one RNNs with $N=40$ neurons on delayed addition tasks ($5000$ epochs, $\tau = 10ms$, $\Delta t = 5ms$, $T=200ms$). 3 out of 100 networks failed to achieve accuracies beyond $0.3$. \textbf{A-B} The learning process across epochs for an example network. \textbf{A} Similar to full-rank RNNs, the learning in rank-one RNN is divided into search and comprehension phases. \textbf{B} During the training, we computed the latent circuit equations for the rank-one RNN, which allowed us to compute the number of fixed points in the network and the flatness score for each epoch. As training progresses, the network learns to build an approximate line attractor between $\kappa \in [0,2]$, which has a non-zero but almost negligible time derivative $\dot \kappa \approx 0$. During training, the latent variable of this example RNN does not undergo any bifurcation. \textbf{C} Using all 97 networks, we studied the number of fixed points, the flatness scores, and the test accuracies as functions of number of epochs ($\Delta \text{Epochs} = 0$ corresponds to the end of the search phase). At the end of the search phase, all quantities showed sharp increases. Solid lines: means. Error bars: s.e.m. over 97 networks. \textbf{D} We computed the flatness scores and the number of fixed points vs the test accuracies of $97$ networks. Flatness scores, but not the fixed points number, showed strong associations with the fraction of correct trials in a typical network. Each dot: median across $97$ networks for a given epoch.}
    \label{fig:fig3}
\end{figure*}

To study the mechanisms behind abrupt skill acquisition, we trained rank one RNNs to perform delayed addition tasks using $\kappa(t)$ as the output of this architecture class (Fig. \ref{fig:fig3}). Similar to their full-rank counterparts, rank one RNNs also showed two phases of learning (Fig. \ref{fig:fig3}\textbf{A-C}), and solved the task by creating ``approximate line attractors'' covering the full output range from $0$ to $2$ (See Fig. \ref{fig:fig3}\textbf{B} for an example, red line). These attractive structures do not need to be exact, rather, a small derivative compared to trial length $T$ ($\dot \kappa(t) \ll T^{-1}$) is sufficient to create an ``approximate line attractor.'' In such a case, any latent state initialized with the cue would reliably be stored, as changes in the latent variable would be negligible following $\Delta \kappa \sim O(\dot \kappa(t) T) \approx 0$. In other words, the networks could use these attractors to store memories to solve the task, as was the case in the prior work \cite{schmidt2021identifying}. We conjectured that the emergence of this structure during learning was the facilitator of abrupt skill acquisition.

First, to gain intuition, we considered an example network in Figs. \ref{fig:fig3}\textbf{A-B}. We extracted the latent circuits at epochs $1000$ (right before the jump in the loss function), $1100$ (during the jump), and $1200$ (right after the jump). Notably, the creation of a flat region approximating a line attractor, which happened in this case without any accompanying bifurcation, led to the jump in the loss function, \textit{i.e.}, what we characterized as skill acquisition in the full-rank RNNs (Figs. \ref{fig:fig1}, \ref{fig:fig2}, and \ref{fig:figs1}). Next, in order to quantify these observations, we trained a total of $100$ networks. Out of all networks, only $3$ failed, while $97$ networks learned to solve the task. We based our analysis in the rest of this section on these $97$ networks. To test our conjecture explicitly, we developed a flatness score (\textbf{Methods}), which measures the total length of $\kappa$ intervals for which $\dot \kappa$ and $\partial_\kappa \dot \kappa$ are both smaller than a user-defined threshold. If the emergent slow points constitute a continuous flat region, they approximate a line attractor, and therefore would have a high flatness score. 

We studied the association between these flatness scores and the test accuracies. First, the flatness scores exhibited a characteristic sudden jump at the end of the search phase, similar to the test accuracies (Fig. \ref{fig:fig3}\textbf{C}). Second, plotting the flatness score as a function of test accuracy (Fig. \ref{fig:fig3}\textbf{D}) revealed a clear association: increased flatness corresponded to better test accuracy. Most importantly, the gradual improvement in test accuracy was accompanied by a steady increase in flatness score (Fig. \ref{fig:fig3}\textbf{D}), which we also confirmed with visual inspection of individual training instances (data not shown). Finally, the average Spearman's rank correlation coefficient between the flatness scores and the relative epochs to the GR events was $0.99 \pm 0.01$ (mean $\pm$ std), and that between test accuracies and the flatness scores was $0.87 \pm 0.08$ (mean $\pm$ std). Together, these observations suggest that the widths of the flat regions gradually increased over time and that small expansions in these flat regions not only reduced the training loss but also were immediately followed by further expansions.

Finally, following prior line of research \cite{eisenmann2023bifurcations,Monfared2020,delmastro2023dynamics}, we asked whether bifurcations played a key role in this emergence of new computational structures (here, approximate line attractors). Fortunately, since $\kappa$ is a one-dimensional dynamical system, we can study all relevant bifurcations as changes in the number of fixed points. (This method can reveal all but one type of bifurcations in one-dimensional dynamical systems, referred to as transcritical bifurcations \cite{strogatz2018nonlinear}, which are nonetheless not directly relevant for the creation of \emph{new} slow regions.) On the one hand, the mean number of fixed points had the distinctive jump at the end of the search phase (Fig. \ref{fig:fig3}\textbf{C}). On the other hand, roughly one fifth of the learned networks had only one fixed point during and after the growth, and thereby did not undergo any bifurcations (including transcritical bifurcations as they require at least two fixed points; data not shown). Furthermore, plotting the (median) number of fixed points as a function of (median) test accuracy did not reveal any clear trend (Fig. \ref{fig:fig3}\textbf{D}). In fact, this plot showed that at least half of the well-trained networks, \textit{i.e.}, the median across all networks, had only one fixed point at the end of the training (Fig. \ref{fig:fig3}\textbf{D}). Therefore, we concluded that skill acquisition in this task could happen without bifurcations, which are still correlated with, but necessary for, the emergence of the approximate line attractors. Instead, the latent circuit can remain close to the bifurcation boundary, but still robustly encode memory despite constant noisy inputs.

\subsection{Characterizing learning dynamics in search and comprehension phases} \label{sec:two_phases}

\begin{figure*}
    \centering
    \includegraphics[width = .9\textwidth]{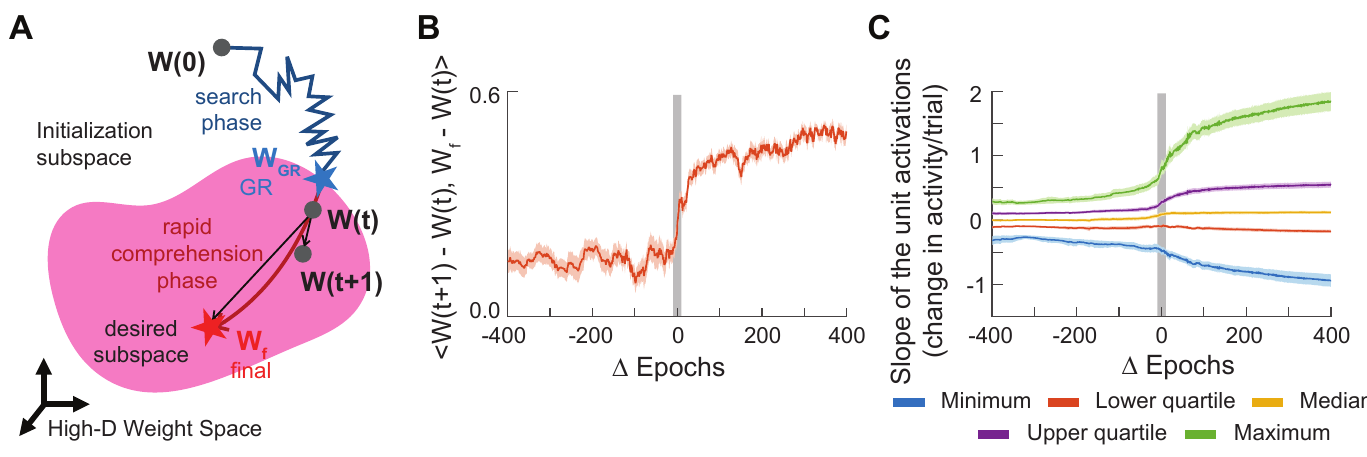}
    \caption{\textbf{Geometric restructuring leads to qualitative changes in the trained networks.} \textbf{A} Schematic of the (conjectured) training process as a dynamical trajectory in the weight space. $W_{\rm GR}$ corresponds to the network parameters at the boundary of the (final) explicit GR event (Figure \ref{fig:figs1}\textbf{C-D}). \textbf{B-C} To study the effects of GR events on the networks, we trained 20 networks with $N=40$ neurons on the delayed addition task ($3000$ epochs, $T=40$). All but one successfully solved the task. \textbf{B} We computed the cosine similarity between the instantaneous update and the optimal update directions, which is used to measure the quality of the learning signal at a given epoch. \textbf{C} The slope of the neural activities, calculated as the change between the activities at the end and the start of the trials, as a function of training epochs. In both plots, $\Delta \rm Epochs=0$ corresponds to the time of abrupt increase in the accuracy, where the search phase concludes and the comprehension phase begins. Lines: means. Shaded regions: s.e.m. over 19 networks.}
    \label{fig:fig4}
\end{figure*}

In this section, we return to (full-rank) PLRNNs and investigate the differences in the learning speeds between the search and comprehension phases (Fig. \ref{fig:fig1}\textbf{B}). We first set out to define a metric that quantifies the quality of learning at a given epoch and study its behavior during the skill acquisition. To achieve this, we analyzed the weight changes in PLRNNs at each training step. Specifically, we measured the alignment of the instantaneous weight change, $W(t+1) - W(t)$, with the optimal learning direction, $W_f - W(t)$, where $W_f$ represents the final weights of the fully trained network (Fig. \ref{fig:fig4}\textbf{A}). We computed this alignment, quantified via the cosine similarity, and plotted as a function of the number of training epochs in Fig. \ref{fig:fig4}\textbf{B}. The learning signal quality significantly improved after the GR event, as demonstrated by comparing the post-event period (averaged within $[+50,+100]$ epochs) with the pre-event period (averaged within $[-100,-50]$ epochs), where epoch $0$ marks the time of the GR event (Wilcoxon signed-rank test, $n=19$ networks, $p<10^{-3}$; \textbf{Methods}).

Although the quality of the learning signal was poor earlier in the training, it was above chance level in both training phases (data not shown). In other words, Adam-based gradient descent was beneficial both before and during the GR event. However, the low gradient alignment during the search phase implies a form of suboptimality in the learning signal (Fig. \ref{fig:fig4}\textbf{B}). In contrast, the quality of the learning signal was much higher in the comprehension phase (Fig. \ref{fig:fig4}\textbf{B}), implying that the gradient is effective in refining the task subserving structures once a desired geometry is roughly present (Fig. \ref{fig:fig3}\textbf{B}). Taken together, these facts tie back to our question at the end of Section \ref{sec:sec2p1} and provide an explanation for the different training speeds observed in the search versus comprehension phases: Gradient is an effective polisher of existing geometries, but not necessarily an optimal facilitator of necessary geometrical restructuring.

\subsection{Empirical predictors of the geometric restructuring events}

To improve the learning signal during the search phase, we next investigated which factors predicted the onset of the GR event. As a first step, we tested different values of weight decay and training batch sizes as potential predictors of shorter search phases. Order-of-magnitude variations in weight decay had little to no effects on the duration of the search phase (Fig. \ref{fig:figs2}), but suboptimal weight decay values led to decreased final performances of the learned networks (Fig. \ref{fig:figs2}). Therefore, even though it was an important parameter to optimize, weight decay did not provide an actionable insight to further probe the learning signal quality in the search phase. In contrast, when we trained PLRNNs with different numbers of batch sizes, we observed a clear trend of decreased search phase with lower batch sizes (Fig. \ref{fig:figs3}). However, randomly perturbing the synaptic weights just before the end of the search phase in a few example networks—without the aid of the gradient—did not result in GR events (data not shown), suggesting a more nuanced relationship between stochasticity and learning than can be leveraged through simple noise addition. We leave it to future work to explore this direction further.

The most consistent and effective predictor we observed was the emergence of ramping  neurons, defined as neurons with slowly (yet monotonically) increasing or decreasing neural activations within each trial. We quantified the emergence of such units by considering the average slopes of neurons as a function of time into the trials (Fig. \ref{fig:fig4}\textbf{C}; \textbf{Methods}). Both the slopes and the number of ramping neurons increased after the geometric restructuring (we visually confirmed that this increase was not due to any oscillatory behavior; data not shown). This observation aligned with our expectations, as neurons with steadily increasing or decreasing activity have been known to be associated with short-term memory processes and timekeeping \cite{schmidt2021identifying, cueva2020low, narayanan2016ramping}. Therefore, a reasonable remedy to the poor learning signal during the search phase may be to design a learning algorithm that incentivizes the emergence of these neurons. However, first, we need to check whether this relationship is correlational or causal in nature.

To test this, we investigated the emergence of slowly varying ramping units in greater detail by studying a recently proposed manifold attractor regularization (MAR) \cite{schmidt2021identifying}. Recent work has developed MAR by incentivizing a subset of neurons to have prolonged time-scales, leading to the formation of ``manifold attractors.'' For the network introduced in Eq. \eqref{eq:ds}, this can be achieved by enforcing $A \approx 1$, $h\approx0$, and $W \approx 0$ for a subset of neurons. In the limit of infinite regularization, this procedure effectively creates a manifold attractor, inside which the neural activities do not change at all unless an outside input perturbs the system. However, in practice, since the regularization is accompanied by a loss term, this manifold is only approximate, \textit{i.e.}, $\dot x(t)$ are not exactly zero, rather vary slowly. As we have discussed in Fig. \ref{fig:fig3}, these approximate manifolds (\textit{i.e.}, generalization of approximate line attractors) can be beneficial for solving the delayed addition task \cite{schmidt2021identifying}. In fact, if we initialized the networks with a manifold attractor initialization (MAI; $A=1$, $W=0$, and $h=0$ for a subset of neurons), the resulting network may not require GR events to create new attractors in the phase space.

As expected, we observed that designing a manifold attractor (by hand) in a subset of units abolished the search phase for the delayed addition task (Figs. \ref{fig:figs4} and \ref{fig:figs5}). As shown in Fig. \ref{fig:figs4}, the units in this subset, termed the ``memory neurons'' (the rest termed the ``computation neurons'') \cite{schmidt2021identifying}, had the same ramping properties that we observed in Fig. \ref{fig:fig4}\textbf{C}. In other words, we discovered that enforcing MAR, which effectively led to slowly varying units, also led to the emergence of ramping units. We hypothesized that networks may use these approximate attractors, and the ramping units, to keep the memory of the two cues. To test this, we decoded the memory of cues in PLRNNs from both memory and computation neuron subpopulations (Fig. \ref{fig:figs5}\textbf{A-C}). As expected, only the former group kept the memory of the two cues throughout the full trial. Overall, we concluded that enforcing the MAR/MAI led to the emergence of slow units, and these memory neurons were actively maintaining the memory of the two cues in the delayed addition tasks. 

Most interestingly, we observed that enforcing MAR (with no MAI) promoted faster training by allowing the network to undergo a GR event closer to network initialization (Fig. \ref{fig:figs5}\textbf{D-F}). Thus, our findings not only confirmed the prior literature (\cite{schmidt2021identifying}), but also led to a testable prediction: Enforcing slow units may accelerate the search phase by facilitating restructuring toward the desired geometrical landscapes. In the next section, we explicitly tested this hypothesis, which eventually led to a network architecture-agnostic form of temporal consistency regularization.

\begin{figure*}
    \centering
    \includegraphics[width = .9\textwidth]{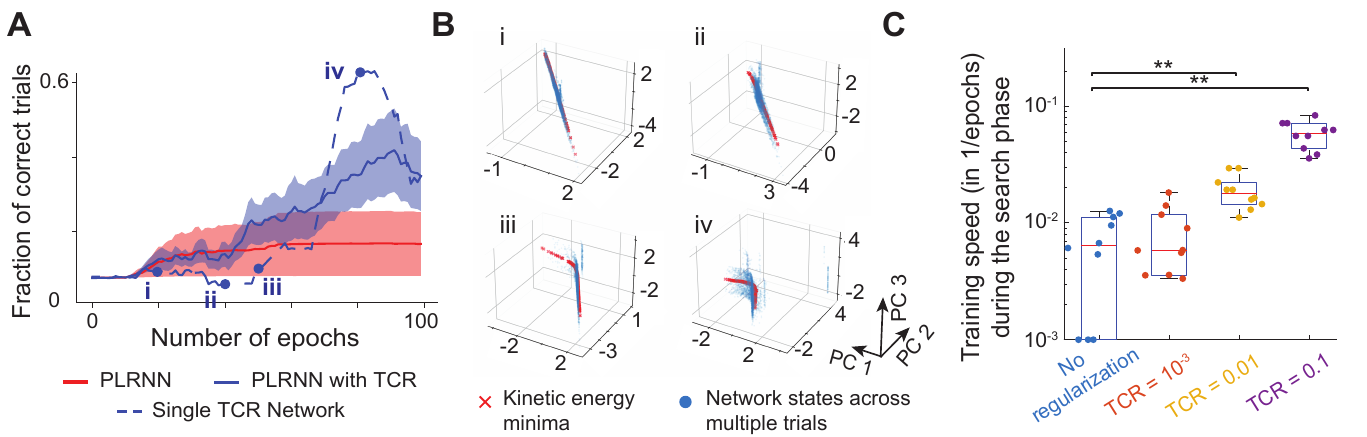}
    \caption{\textbf{Temporal consistency guides the learning toward parameter regions with favorable geometries.} We trained 10 regularized and unregularized PLRNNs with $N=40$ neurons on the delayed addition task (500 epochs, $T=40$). For regularized networks, $N_{\rm reg}=20$ out of $N=40$ were memory neurons with $\lambda_{\rm TCR} = 1$. \textbf{A} Enforcing temporal consistency in a subset of neurons, and thereby explicitly encouraging the formation of slow units, increased the number of networks passing the search phase and undergoing the abrupt skill acquisition. The dotted line corresponds to an example network. Solid lines: means. Shaded areas: s.e.m. over 10 networks. \textbf{B} We plotted the attractor landscape of an example network, corresponding to the (i) - (iv) stages shown in \textbf{(A)}. The network underwent geometric restructuring, in which a second arm was created on the slow point landscape (red crosses). \textbf{C} Increasing the TCR regularization strength led to a faster training process during the search phase. Comparisons are performed over 10 networks with two-sided Wilcoxon signed-rank tests ($^{**}p<10^{-2}$).}
    \label{fig:fig5}
\end{figure*}

\subsection{Temporal consistency incentivizes abrupt formation of attractors}
So far, we have observed that the emergence of slow units may increase memory capabilities, which can be explicitly enforced to improve training efficiency in PLRNNs performing the delayed addition task (Fig. \ref{fig:figs5}, \cite{schmidt2021identifying}). Inspired by this observation, we now propose the temporal consistency regularization (TCR), a (potentially) biologically plausible mechanism that is agnostic to the network architecture and the details of the desired geometric landscapes:
\begin{equation}
    \mathcal{L}_{\text{TCR}}= \frac{1}{T} \sum_{t=1}^T \sum_{i=1}^{N_{\text{reg}}} (x_i[t] - x_i[t-1])^2.
\end{equation}
Intuitively, TCR encourages slow time dynamics for a subset ($N_{\text{reg}}$ out of $N$) of neurons, which we term as ``memory neurons.''  Then, regularizing task training with TCR, \textit{i.e.}, $\mathcal{L} = \mathcal{L}_{\rm task} + \lambda_{\rm TCR} \mathcal{L}_{\text{TCR}}$, may incentivize attractor formation and lead to increased short-term memory capabilities, which we explicitly tested next.

As a first step, we once again considered the PLRNNs in Eq. (\ref{eq:ds}), which were trained to perform delayed addition tasks (Fig. \ref{fig:fig5}). Compared to unregularized networks, training PLRNNs with TCR led to shortened search phase and faster convergence (Fig. \ref{fig:fig5}\textbf{A}). Using the energy minimization analysis (following Eq. (\ref{eq:slow_points})), we further confirmed that the search phase was indeed concluded with a GR event (Fig. \ref{fig:fig5}\textbf{B}). Furthermore, higher values of $\lambda_{TCR}$ resulted in faster training during the search phase (Figs. \ref{fig:fig5}\textbf{C} and \ref{fig:figs6}), signaling the direct involvement of TCR in shortening the search phase. As a second step, we asked whether TCR can achieve abrupt convergence on par with enforcing optimal solutions. Given that MAR establishes a plane attractor in the network (\textbf{Methods}), a feasible solution for the delayed addition task, we can consider the search phase durations of networks with MAR as an upper/desirable baseline. As expected, the convergence speeds with TCR were generally slower. However, the difference decreased with increasing values of $\lambda_{\rm TCR}$ (Fig. \ref{fig:figs6}). Hence, TCR (similar to MAI/MAR) incentivized attractor formation and led to faster search phases in PLRNNs trained to perform delayed addition tasks.

\begin{figure*}
        \centering
    \includegraphics[width = \textwidth]{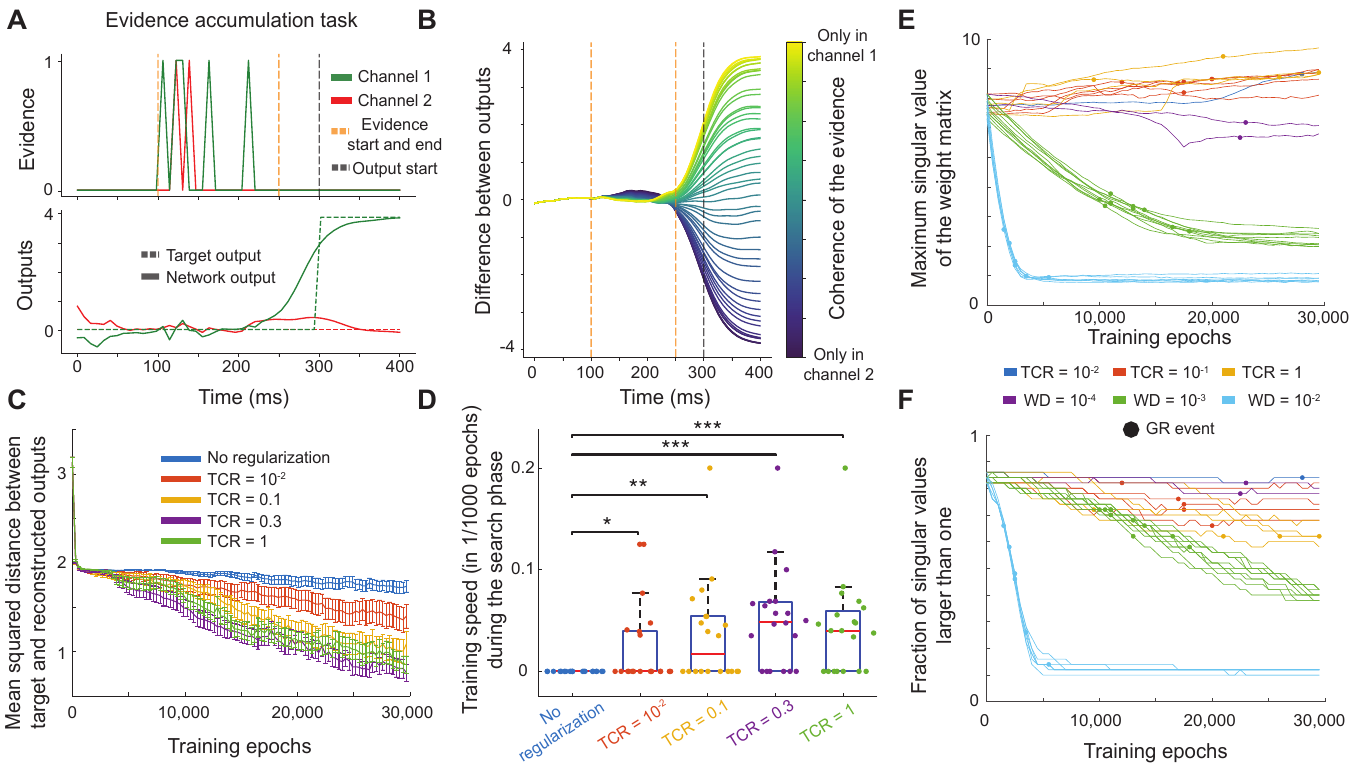}
    \caption{\textbf{Temporal consistency allows training strongly connected RNNs to perform the evidence accumulation task.} \textbf{A} An illustration of the evidence accumulation task with two input channels. The network receives binary pulses from both channels, has to decide which channel had more ``evidence," and outputs its identity. \textbf{B} The differences between network outputs are modulated based on the coherence of the evidence. A lower coherence leads to less confident and sometimes inaccurate network predictions, whereas the network is almost always correct in high coherence trials. \textbf{C-D} We trained 20 RNNs ($N=50$, $N_{\rm reg} = 25$, $\tau = 10ms$, $\Delta t = 8ms$, and $g = 4$) for $30,000$ epochs to solve the evidence accumulation task. \textbf{C} The evolution of the mean squared error between the (scaled) coherence of trials and the differences in output channels plotted as a function of training epochs. Networks without TCR were not able to solve the task. Solid lines: means. Error bars: s.e.m. over 20 networks. \textbf{D} The training speed during the search phase as a function of temporal consistency regularization strength. Only regularized networks were able to pass into the comprehension phase. The comparisons were performed with two-sided Wilcoxon signed-rank tests ($^{*}p<0.05$,$^{**}p<10^{-2}$, and $^{***}p<10^{-3}$). \textbf{E-F} Finally, we performed control experiments using the same setup as in \textbf{C-D}, but also using weight decay values. Though weight decay led to more networks concluding the search phase, this was often accompanied by decreases in the singular values of the weight matrix (\textbf{E}) and in the number of singular values that were larger than one (\textbf{F}). Each solid line corresponds to an RNN that had successfully passed the search phase. The dots correspond to the timings of the GR events. For panels \textbf{(E-F)}, we trained 10 networks per condition.}
    \label{fig:fig6}
\end{figure*}

\subsection{TCR enables abrupt learning in strongly connected recurrent neural networks}

To illustrate the network- and task-agnostic nature of TCR, we next focused on leaky firing rate RNNs (lfRNNs) \cite{dinc2023cornn,masse2019circuit}:
\begin{equation}
\label{eq:lfr}
    \tau \frac{\diff r(t)}{\diff t}= -r(t) + \tanh(Wr(t) + Cs(t) + h). 
\end{equation}
Here, $\tau \in \mathbb R$ is the predefined, non-trainable, neuronal decay time, other components are defined similarly to Eq. (\ref{eq:ds}), and with $x(t) = Wr(t) + Cs(t) + h$. For our purposes, we set $h = 0$, as our tasks of interest can be trained without bias. In practice, we discretize Eq. (\ref{eq:lfr}) with a time step $\Delta t$ (\textbf{Methods}). 

Next, we trained lfRNNs to perform evidence accumulation tasks (Figs. \ref{fig:fig6} and \ref{fig:figs7}\textbf{A-B}). In an evidence accumulation task, there are two input and two output channels to the network (Fig. \ref{fig:fig6}\textbf{A}). The input pulses in the input channels are sampled from a binomial distribution for each time point, and the network is tasked to indicate which input channel received more pulses, \textit{i.e.}, evidence, by outputting a pulse in the respective output channel. Higher coherence of the evidence, \textit{i.e.}, when higher fraction of the total evidence was presented predominantly in one of the channels, corresponds to an easier task (Fig. \ref{fig:fig6}\textbf{B}).

\begin{figure*}
    \centering
    \includegraphics[width=0.9\textwidth]{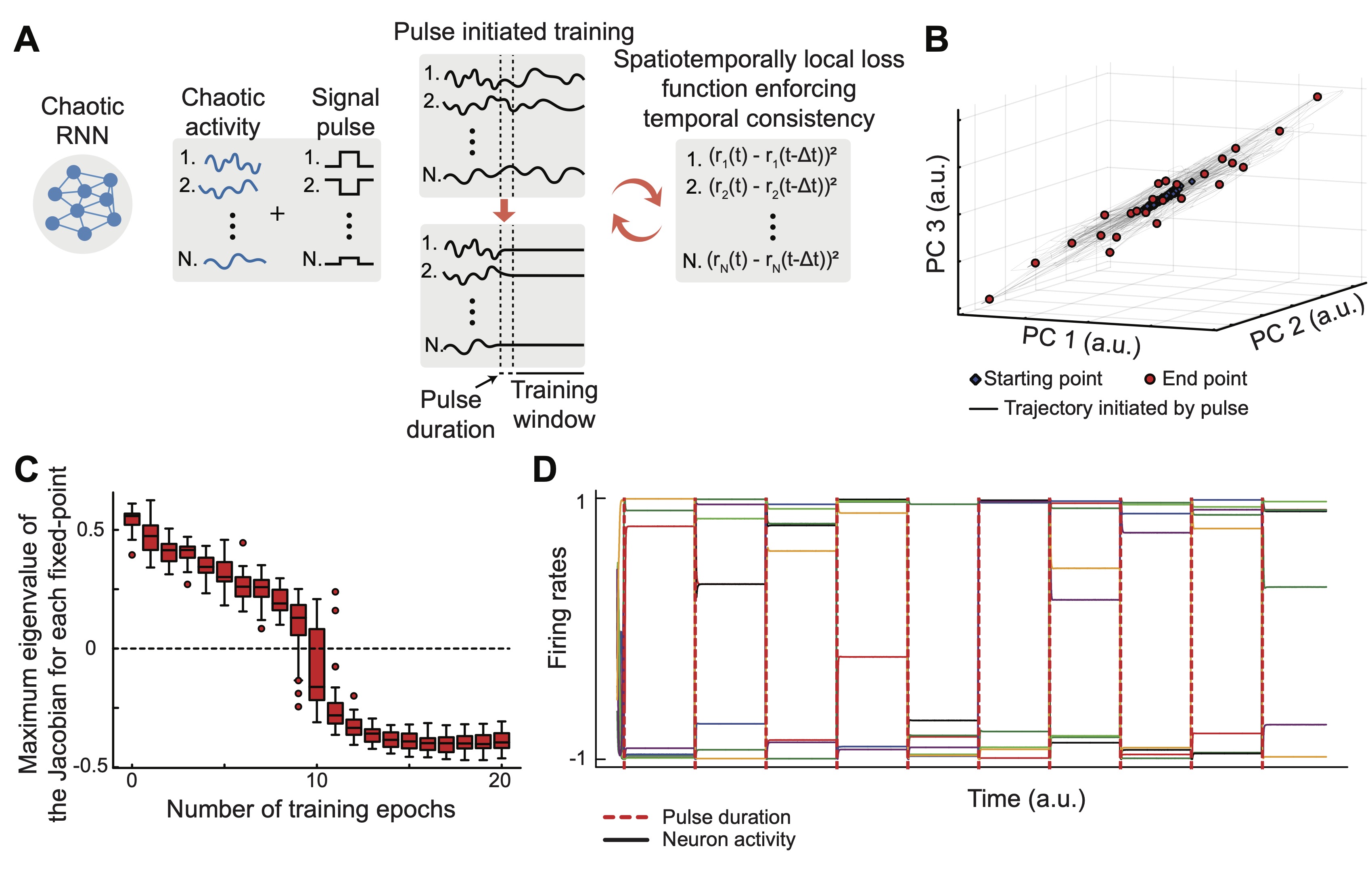}
    \caption{\textbf{Temporal consistency facilitates online learning of cue-responsive fixed points.} \textbf{A} The training procedure starts with a randomly initialized RNN. A rectangular pulse is applied through a randomly initialized, non-trainable input weight matrix to the network. For a pre-defined duration after the pulse offset, the freeze-in training takes place. \textbf{B} We visualized the final output of a fully trained network with $25$ distinct, cue-responsive fixed point attractors. \textbf{C} The maximum eigenvalues of the Jacobian matrices became non-positive (attractive fixed points) after as little as $12$ epochs of training. \textbf{D} We visualized $10$ out of $25$ cue-responsive fixed points. Note that after the offsets of the input pulses, which brings the network nearby but not exactly to its final destination, the network briefly evolves and settles in the cue-responsive fixed points. Pulses are shown during a very short duration (red dotted lines), and drive the transitions between fixed points.}
    \label{fig:fig7}
\end{figure*}

As a first test, we confirmed that even unregularized lfRNNs can solve this task when initialized with Xavier initialization, although TCR still increased the speed with which the search phase was concluded (Fig. \ref{fig:figs7}\textbf{A-B}). Surprisingly, TCR showed its true value when lfRNNs were initialized in a strongly connected regime (controlled by a parameter, $g$, see \textbf{Methods}), where recurrent connections are notoriously difficult to train \cite{sussillo2009generating,depasquale2018full,maass2002real,jaeger2004harnessing}. Specifically, even after 30,000 training epochs, none of the $20$ unregularized networks was able to pass the search phase (Fig. \ref{fig:fig6}\textbf{C}). For these networks, the errors plateaued at a particular value, similar to the plateaus of the loss curve of the delayed addition task (Figs. \ref{fig:fig1} and \ref{fig:fig2}). This was only overcome by networks trained with TCR (applied to $N_{\rm reg} = 25$ out of $N=50$ neurons), where $14$ out of $20$ networks learned to solve the task. Furthermore, increasing the TCR strength, $\lambda_{\rm TCR}$, increased the training speed (until saturated) during the search phase (Fig. \ref{fig:fig6}\textbf{D}), similar to the delayed addition task (Fig. \ref{fig:fig5}\textbf{C}). Notably, weight decay also led to improved training for these RNNs, though this was accompanied by substantial decreases in the singular values of the weight matrix (Fig. \ref{fig:fig6}\textbf{E-F}), \textit{i.e.}, likely by taking the network out of the strong connection region. Although TCR was less intrusive in the global singular value structure (Fig. \ref{fig:fig6}\textbf{E-F}), we leave it to future work to study in depth the chaotic (or lack thereof) nature of solutions found by TCR in general settings.

Finally, we provided additional evidence for the generality of our findings so far with two experiments: i) we trained strongly connected lfRNNs on 3-bit flip flop tasks (Fig. \ref{fig:figs7}\textbf{C-E}), and ii) we trained long short-term memory (LSTM) networks on delayed addition tasks (Fig. \ref{fig:figs7}\textbf{F-H}). In both cases, in line with our results so far, TCR led to faster training during the search phases.

\subsection{Cue-responsive fixed points with online freeze-in learning}

In essence, TCR forces the derivative of network activities towards zero for a subset of neurons. Discretization of this term leads to an error signal that is simply the difference between current and previous neural activities, which can be applied as a local and online (potentially biologically plausible) learning rule (\textbf{Methods}): 
\begin{equation} \label{eq:online}
\Delta W_{ij}[t] = - \lambda_{\rm TCR} \left(r_i[t]-r_i[t-\Delta t]\right)r_j[t-\Delta t],
\end{equation}
where $\Delta W[t]$ stands for the changes in the recurrent weights. The update rule $\Delta W_{ij}$ operates locally, depending only on the activities of the connected pre-synaptic ($j$) and post-synaptic ($i$) neurons. Its online nature stems from using just two immediate signals: the derivative of the post-synaptic neuron's activity and the pre-synaptic neuron's most recent activity. Such a mechanism could be biologically plausible, provided future research discovers plasticity processes capable of tracking neural activity derivatives.

Building on the foundational work of Hopfield networks and associative memories \cite{hopfield1982neural}, we applied this learning rule to create cue-responsive fixed points (Fig. \ref{fig:fig7}). Specifically, we trained a network such that each fixed point responds to a specific cue, \textit{i.e.}, when the cue is presented, it drives the network state toward the vicinity of its associated fixed point. The network's intrinsic dynamics then naturally guide the state to converge to that fixed point, hence the name ``cue-responsive fixed points." While traditional Hopfield networks establish fixed points through carefully constructed weight matrices that encode desired memory patterns \cite{hopfield1982neural}, our approach revealed a striking phenomenon: these fixed points can emerge abruptly through online training in a self-organizing manner (Fig. \ref{fig:fig7}\textbf{A}). 

To establish these fixed points, we sequentially presented distinct input cues to the network through separate input channels. During each presentation, we updated the network weights using the online learning rule described in Eq. \eqref{eq:online} - a process we termed ``freeze-in training.'' This approach successfully created 25 distinct fixed point attractors (by choice, not due to some inherent limit), each associated with its corresponding cue (Fig. \ref{fig:fig7}\textbf{B}). The training process proved remarkably efficient, rapidly establishing attractive dynamics around each fixed point in ass little as $12$ presentation of the cue sequence (Fig. \ref{fig:fig7}\textbf{C}). Importantly, while cue inputs could drive the network to these fixed points, the points' stability persisted long after the cue offset, demonstrating that their existence was independent of the input signals (Fig. \ref{fig:fig7}\textbf{D}). The cues were simply associated with them.

Overall, using an online version of temporal consistency, we successfully trained networks to develop cue-responsive fixed point attractors without requiring global learning signals or explicit supervision. This represents a notable advance over traditional approaches like Hopfield networks \cite{hopfield1982neural}, which required careful weight initialization and operated under strict constraints, and task-based RNN training methods \cite{sussillo2013opening}, which relied on external supervision signals. The freeze-in training, instead, suggests a novel approach for designing networks with associative memories, in which these computational building blocks can emerge naturally through online learning in a (potentially) biologically-plausible manner.

\section{Discussion}

It is not surprising that (approximate) attractors would need to be formed to enable new computational capabilities, and existing work often focused on bifurcations as the main driver of such mechanisms \cite{eisenmann2023bifurcations,pmlr-v202-hess23a,delmastro2023dynamics}. Interestingly, despite key insights from dynamical system theory stating that such bifurcations should be desirable to form attractor landscapes, a common theme in learning dynamics of artificial neural networks relates to how to suppress bifurcations, as they are often associated with vanishing/exploding gradient problems \cite{doya1992bifurcations,pascanu2013difficulty}. In fact, even recent seminal work focusing on dynamical system theory discusses certain bifurcations that need to be avoided and provides guidance on how to avoid them \cite{eisenmann2023bifurcations}. Thus, learning in recurrent systems has to deal with a clear paradox: dynamical system theory suggests that bifurcations are beneficial for learning new attractor structures \cite{strogatz2018nonlinear}, yet analysis of learning dynamics suggests that these abrupt changes lead to destabilization of gradient dynamics \cite{doya1992bifurcations}. This raises two important questions: how do we reconcile these seemingly contradictory statements, and is this focus on bifurcations stemming from the practicality associated with their analysis or are they the sole mechanisms of endowing networks with novel computational capabilities? 

In this work, we showed that dynamical system theory provides a negative answer on the latter: we demonstrated that a more general set of mechanisms plays a role in both making learning dynamics nontrivial and endowing the networks with new capabilities. We achieved this by taking a bifurcation-agnostic approach to studying the abrupt changes to the networks' computational capabilities and coined the term ``GR events'' to describe this broader set of qualitative changes in the RNNs' phase spaces. In doing so, we provided an answer to the first question: although gradients destabilize during GR events, such events nonetheless are eventually beneficial to the training of RNNs and incentivizing them facilitates fast learning of desirable attractor landscapes.

\subsection{Learning signal quality before and after auto-catalytic growth}

Our analysis reveals a key insight about networks initialized far from the desired geometry: they may exhibit sustained epochs of apparent non-learning, yet experience regions of gradient instability that signal qualitative changes in the slow point landscape, indicating hidden GR events (Fig. \ref{fig:fig2}). Rather than being obstacles to be reverted, these instability regions may mark transitions after which the network continues training with a new geometry. In the delayed addition task, we observed that such transitions led to an abrupt growth of approximate line attractors, ultimately enabling short-term memory computation. This growth manifested itself in the loss function as spikes or sustained jumps. We found similar gradual yet rapid behaviors across various tasks and architectures, contrasting with traditional bifurcations where attractors form immediately after the final bifurcation. Instead, the formation of a small portion of the desired phase space structure triggers (abrupt) growth, enabling new capabilities without bifurcations. During this growth process, we observed improved gradient signal learning quality compared to earlier epochs. This suggests that learning difficulties and prolonged loss plateaus cannot be attributed solely to vanishing gradients, as is the traditional wisdom \cite{pascanu2013difficulty,ribeiro2020beyond}. Rather, the gradient \emph{direction} itself is misaligned during the search phase. Consequently, addressing vanishing gradients alone would not resolve grokking-like behavior in RNNs trained for short-term memory tasks.

Our findings support the generalization of insights from low-dimensional dynamical systems \cite{strogatz2018nonlinear}: the weight space comprises multiple subspaces with distinct geometries of the attractor structure (Fig. \ref{fig:fig3}). We discovered that gradient updates effectively drive the network toward optimal weights only when the network already supports the appropriate computational structures. Before reaching this optimal weight subspace, the network must traverse one or more intermediate regions with minimal, though non-zero, gradient signal assistance. This suggests that accelerating training in large-scale foundational models may require not just new optimizers \cite{adamOpt_kingma2017adam}, but loss functions whose gradients incentivize beneficial GR events and minimize the time spent in the search phase.

\subsection{Improving learning dynamics by explicitly incentivizing GR events}

While analyzing GR events in task-trained RNNs, we provided the first account of \emph{incentivizing} beneficial GR events to facilitate faster learning in short-term memory tasks. Although we found that gradients destabilize during GR events (Figs. \ref{fig:fig2} and \ref{fig:figs1}\textbf{C-D}), this destabilization may be advantageous. Specifically, we observed poor learning signal quality before the GR events (Fig. \ref{fig:fig4}), suggesting that learning in RNNs benefits from these destabilizing regions. Additionally, our re-examination of manifold attractor regularization for PLRNNs \cite{schmidt2021identifying} showed that enforcing slow-time dynamics can induce attractor formation, enhancing short-term memory capabilities. These insights led to our development of \textit{temporal consistency regularization}, an attractor-agnostic, architecture-agnostic, and biologically plausible mechanism to induce GR events.

In this work, we assigned a subset of neurons the role of ``memory neurons,'' which were encouraged to have slow-time dynamics (Fig. \ref{fig:fig5}). The resulting temporal consistency regularization allowed training leaky firing rate RNNs, which do not have a RELU non-linearity or trainable neuronal time scales, under even strongly connected initialization and potentially in an online manner (Figs. \ref{fig:fig6} and \ref{fig:fig7}). Unlike prior solutions to training strongly connected RNNs \cite{sussillo2009generating}, we utilized temporal consistency regularization in conjunction with back-propagation through time, which allowed using established algorithms (here, Adam optimizer \cite{adamOpt_kingma2017adam}) for training RNNs. 

Similar forms of temporal-consistency loss have been utilized in computer vision research \cite{TC_guan2021domain, TC_Park2019, TC_chen2017coherent, TC_Huang2017, TC_Ruder2016}. Similarly, \cite{kelly2020learning} has argued that enforcing smaller first derivatives could allow taking large steps to simulate dynamical systems faster. Finally, most relevantly, \cite{sussillo2015neural} has utilized a penalty term that regularized the first-order dynamics in RNNs to incentivize RNNs to find simpler final solutions. Our work represents the first application of the temporal consistency regularization in the context of dynamical system theory, specifically to incentivize GR events and to induce efficient learning dynamics in RNNs.

\subsection{A learning mechanism for training strongly connected networks}

When their weights are sampled from a zero-mean Gaussian distribution with a standard deviation beyond some critical values, RNNs are known to show chaotic behavior \cite{sompolinsky1988chaos}. Consequently, training such RNNs has been a major source of controversy in the last few decades \cite{sussillo2009generating,depasquale2018full,maass2002real,jaeger2004harnessing}. Research in the field often uses non-chaotic initialization schemes such as Xavier initialization \cite{glorot2010understanding}. Yet, understanding how to train chaotic networks is fundamental to our studies of brain dynamics, which regularly show sustained, non-vanishing, neural activities \cite{mattera2024chaotic}. Earlier research in this direction often took heuristic approaches. Most notably, the reservoir computing paradigm proposed the idea that recurrent weights are not trained and that instead one trains only a readout vector to the output, which is then fed back to the RNN \cite{maass2002real,jaeger2004harnessing}. Later, the FORCE algorithm proposed algorithmic changes to ensure that this training was stabilized \cite{sussillo2009generating}. Latest approaches propose using teacher-student paradigms to train randomly initialized networks to mimic the activities of a target network that can solve the desired task \cite{depasquale2018full}. However, the question of how to train strongly connected RNNs with proper gradient signals has remained underexplored. As noted above, one potential solution is to enforce significant weight decay, which could eventually allow the RNNs to escape the strong connection regime. Here, we proposed an alternative solution without a substantial global decay of weight values: TCR enables training RNNs that are initialized deep into the strongly connected regime. 

\subsection{A potential bioplausible learning rule: Changed rates change synaptic weights}

Our work may explain the mechanism of a widely observed phenomenon in systems neuroscience: slow-time dynamics and ramping neural activities play a vital role in memory and associative computations \cite{schmidt2021identifying,cueva2020low, narayanan2016ramping}. The traditional view is that attractor structures \emph{cause} slow-time dynamics (which is true by the definition of an attractor since $\dot x(t) = 0$). However, as we have shown in this work, another plausible explanation for this phenomenon may include reversing the causality, \textit{i.e.}, directly encouraging slow-time dynamics in biological networks can promote attractor formation, thus shortening the search phase via beneficial bifurcations. Moreover, the temporal consistency may carry long-term dependencies in an online scenario (Fig. \ref{fig:fig7}), without requiring explicit calculation of the gradients of a global loss function, \textit{e.g.}, via backpropagation through time, which often fails in strongly connected networks to begin with \cite{maass2002real,jaeger2004harnessing}. Finally, reminiscent of Hopfield networks \cite{hopfield1982neural}, freeze-in training allows building cue-responsive fixed point attractors by simply slowing down the neural activities for a short period after the observation of a cue, \textit{i.e.}, self-organization without explicit task-training. Consequently, a testable biological prediction of our work is the putative existence of a very specific form of short-term synaptic plasticity mechanism that implements TCR biologically.

\section{Outlook}

Collectively, our work represents a significant advancement in understanding learning mechanisms in biological and artificial networks, highlighting the importance of investigating currently underexplored areas in high-dimensional dynamical system theory. Our approach differs fundamentally from previous work that focused on specific attractor structures in dynamical system reconstruction problems \cite{schmidt2021identifying,eisenmann2023bifurcations}. Instead of assigning particular attractor structures or architectures, we studied short-term memory tasks with sparse learning signals during delay periods, allowing attractor landscapes to emerge naturally through task training. This approach enabled us to extract mechanistic insights into how attractors that support short-term memory emerge and evolve.

Looking ahead, several promising directions follow from our findings. First, while our temporal consistency regularization proved effective, its current implementation using PyTorch's \textit{RNNCell} module introduces significant computational overhead compared to the \textit{RNN} module. The development of specialized modules allowing direct access to internal RNN states could substantially improve training efficiency for more complex applications. Second, our insights into how networks traverse different attractor subspaces suggest potential strategies to accelerate training in large-scale models beyond traditional optimizer improvements. Third, our biological predictions about specific forms of short-term synaptic plasticity that implement TCR could guide future experimental studies in systems neuroscience. Finally, our preliminary studies revealed an intriguing distinction between TCR and weight decay: unlike weight decay, TCR tended to preserve largest singular values in strongly connected RNNs. This observation raises a fundamental question about whether TCR guides networks out of the chaotic regime entirely or instead harnesses (or tames, \textit{e.g.}, \cite{laje2013robust}) chaotic dynamics to solve the task.

Before concluding, it is worth noting that our work is not without its limitations. In our experiments, we observed that TCR is most beneficial if the initialization is far away from the desired weight subspaces, \textit{e.g.}, the strong weight initialization ($g=4$) in Fig. \ref{fig:fig6}. Yet, if the network is already initialized close to the solution or in a non-chaotic regime, the benefits were incremental at best (Fig. \ref{fig:figs7}\textbf{A-B}). Moreover, if the solution requires rapid changes in neural activities (\textit{e.g.}, sin-wave generation), and not formation of memory subserving attractor subspaces, it is possible that TCR may be detrimental. This was particularly true for when RNNs already underwent the GR events, and increased TCR often led to slower learning after the geometric transition (Fig. \ref{fig:figs6}). Finally, another interesting question is to study what stops freeze-in training (Fig. \ref{fig:fig7}) from converging to a trivial fixed point attractor, \textit{i.e.}, $W=0$. It may be desirable to include some random stochastic updates to the recurrent weights, which may prevent such trivial solutions, or design a controlling process for $\lambda_{\rm TCR}(t)$. Thus, the effects of adding noise to the weights and/or neural activities, and the temporal control of the TCR strength constitute limitations of the current work and should be addressed by future work.

Overall, by bridging theoretical insights from dynamical systems with practical challenges in neural network training, our work opens new avenues for understanding and improving learning in both artificial and biological networks. Future work exploring these directions may not only advance our theoretical understanding but also lead to more efficient and robust training methods for complex recurrent neural architectures.

\emph{Acknowledgements} -- We thank Dr. Itamar Landau, Dr. Boris Shraiman, Dr. Surya Ganguli, Mert Yuksekgonul, and Abby Bertics for their helpful feedback on the manuscript and valuable interactions throughout the project. MJS gratefully acknowledges funding from the Simons Collaboration on the Global Brain, the Vannevar Bush Faculty Fellowship Program of the U.S. Department of Defense, and Howard Hughes Medical Institute. FD receives funding from Stanford University's Mind, Brain, Computation and Technology program, which is supported by the Stanford Wu Tsai Neuroscience Institute. NM acknowledges funding from the National Science Foundation, award 2313150. This research was supported in part by grant NSF PHY-2309135 and the Moore Foundation Grant No. 2919.02 to the Kavli Institute for Theoretical Physics (KITP).

\appendix

\renewcommand\thesection{S\arabic{section}}
\renewcommand\thetable{S\arabic{table}}
\renewcommand\theequation{S\arabic{equation}}
\renewcommand\thetheorem{S\arabic{theorem}}
\renewcommand\thelemma{S\arabic{lemma}}
\renewcommand\thecorollary{S\arabic{corollary}}
\renewcommand\thedefinition{S\arabic{definition}}

\setcounter{equation}{0}
\setcounter{theorem}{0}
\setcounter{lemma}{0}
\setcounter{definition}{0}
\setcounter{corollary}{0}
\section{Methods}

\subsection{Network architectures} \label{app:network_architectures}

\subsubsection*{Piece-wise linear recurrent neural networks}
We considered piece-wise linear recurrent networks (PLRNNs)  \cite{koppe2019identifying,schmidt2021identifying} for most of our experiments unless otherwise specified. PLRNN dynamical systems equation is as follows.

\begin{equation}
    x[t]= Ax[t-1] + W\phi(x[t-1]) + Cs[t] + h 
\end{equation}

Here, $x[t] \in \mathbb{R}^{N}$ is the network activations. $N$ number of neurons, $\phi(x[t])$ is the network firing rates with ReLU non-linearity; $\phi(x[t])_i = \text{max}(0, x_i[t]), i \in {1, ..., N}$. $A \in \mathbb{R}^{N \times N}$ is a diagonal matrix encoding decay time constants of neurons. $W \in \mathbb{R}^{N \times N}$ is the recurrent connectivity matrix. $s[t] \in \mathbb{R}^{K}$ is the external input with shape $K$. $C \in \mathbb{R}^{N \times K}$ injects inputs into neurons. $h \in \mathbb{R}^{N}$ is the bias.

To obtain the predictions $\hat{o}[T] \in \mathbb{R}$, we linearly project the activations at the end of the trials $x[T]$.
\begin{equation}
    \hat{o}[T]= W_{\text{out}} x[T] + b_{\text{out}}
\end{equation}
Here, $W_{\text{out}} \in \mathbb{R}^ {1 \times N}$ projects the network activations onto the final predictions. $b_{\text{out}} \in \mathbb{R}$ is the output bias. Unless otherwise specified, we allow the network to learn $A, W, C, h, W_{\text{out}}$ and $b_{\text{out}}$ to minimize the mean squared error between the target ($o_T = u^1_{t_1} + u^1_{t_2}$ for the delayed addition task) and the network predictions $\hat{o}[T]$. 

\subsubsection*{Leaky firing rate recurrent neural networks}
For the analysis on networks with fixed time scales, we used the leaky firing rate RNNs:
\begin{equation}
    \tau \frac{\diff r(t)}{\diff t}= -r(t) + \tanh(Wr(t) + Cs(t)), 
\end{equation}
where $\tau$ is the fixed neuronal time scales and the rest is defined as above. In practice, we used a discretized version of these equations:
\begin{equation}
    r[t+\Delta t] = (1-\alpha)r[t] + \alpha \tanh(Wr[t] + Cs[t]),
\end{equation}
where $\alpha = \Delta t/\tau$ is the discretization constant. We used these equations to perform forward and backwards propagation in all experiments. For all experiments, $\tau = 10ms$ and $\Delta t = 8ms$, leading to $\alpha = 0.8$. When lfRNNs are initialized in chaotic regime, we sampled the weight matrices from a Gaussian distribution with zero mean and standard deviation of $g/\sqrt{N}$ with $g \geq 1$. In these networks, whenever applicable, we enforced temporal consistency on both $r[t]$ and the currents $x[t] = W r[t] + Cs[t]$.

\subsection{Task details}
\label{sec:suppl_tsks}

\subsubsection*{Delayed addition task} 

The delayed addition task consists of several train and test trials. Input to each trial, $U = \{(u^1_1, u^2_1), (u^1_2, u^2_2), ..., (u^1_T, u^2_T)\}$, has the shape of $T \times 2$, where $T$ is the trial length. Here, 
\begin{equation}
  u^1_t  \sim \mathcal{U}[0, 1)   
\end{equation}
is sampled randomly from a uniform distribution between $0$ and $1$, whereas
\begin{equation}
  u^2_t =
    \begin{cases}
      1 & \text{if $t= t_1$ or $t= t_2$}\\
      0 & \text{otherwise}
    \end{cases}       
\end{equation}
is the cue signal that is mostly zero except for times $t=t_1$ and $t=t_2$. Here, $t_1$ and $t_2$ are randomly sampled such that $t_1 < 10$ and $t_2< T/2$. The goal of the task is to output $u^1_{t_1} + u^1_{t_2}$ at the final time step $T$. This requires the network to have a memory that can store a number for at least $T/2$ time steps. To measure the network performances, we calculated the fraction of correct trials. Specifically, we counted the trials where $|o_T - \hat{o}[T]| < 0.04$ and presented it as a fraction. The task was used in previous work to identify memory capabilities in PLRNNs \cite{schmidt2021identifying}.

\subsubsection*{Evidence accumulation task} 

In the evidence accumulation task, the network receives two inputs and has two corresponding outputs. Two inputs contain transient pulses during the evidence interval, at random time points. The task is to provide a non-zero output after a certain time point and only in the output channel corresponding to the input channel with the higher number of transient pulses, \textit{e.g.}, evidence. The network should sustain this output until the end of the trial.  See Fig. \ref{fig:fig6}\textbf{A} for an illustration.

\subsubsection*{3-Bit Flip Flop Task} 

In the 3-bit flip flop task, the RNN has 3 outputs corresponding to the state of 3 memory bits and receives transient $\pm1$ pulses from the corresponding 3 inputs. The task is to output in each channel the latest sign of the corresponding input channel, leading to a total of $2^3$ memory states. The network is required to sustain that value until the corresponding input changes. See Fig. \ref{fig:figs7}\textbf{C} for an illustration.

\subsection{Manifold attractor regularization} \label{app:mar}

Recent work has demonstrated the promise in promoting attractor formation \cite{schmidt2021identifying}. In this framework, memory units are regularized to form a line-attractor subspace, while computation units remain unregularized, resulting in enhanced memory capacity:
\begin{equation}
    \mathcal{L}_{MAR}= \lambda_{\rm MAR} \left[ \sum_{i=1}^{N_{reg}} (A_{ii}- 1)^2 +  \sum_{i=1}^{N_{reg}} \sum_{\substack{j=1\\ j\neq i}}^N W_{i,j}^2 +  \sum_{i=1}^{N_{reg}} h_i^2 \right].
\end{equation}
Here, MAR enforces the diagonal time scales to be $1$ (perfect memory of previous state), and the weights and biases to be zero. If this is exactly enforced on the inputs ($\lambda_{MAR} \to \infty$), we are left with:
\begin{equation}
    x[t+1] = x[t] + Cs[t],
\end{equation}
which is termed a manifold attractor \cite{schmidt2021identifying} and performs integration. In the case of manifold attractor initialization (MAI), only a subset of neurons are initialized with this manifold attractor, creating a sub-circuit capable of performing integration of inputs. 

Overall, the slow time dynamics induced by MAI/MAR directly stem from the fact that they incentivize explicit plane attractor formation, which is specifically beneficial to the delayed-addition tasks. Moreover, MAI/MAR require the dynamical system equation to be in a specific format, making them inapplicable to general architectures.

\subsection{Energy minimization}
\label{app:energy_min}

To extract the slow points of the network, the core implementation is based on previously published work \cite{sussillo2013opening}, though we modified it slightly by replacing the minimization algorithm with a Pytorch based solver to speed up the process (similar to a later implementation \cite{golub2018fixedpointfinder}). We first picked a data batch with 100 trials from the test set. We obtained activations for all the trials. This resulted in $100 \times T$ total number of states. 500 states were randomly picked ($x[0]$) as the initialization points for energy minimization. Starting from each $x[0]$, we minimized energy using stochastic gradient descent in a recursive manner:

\begin{algorithm}[H]
\caption{Energy minimization}
\label{alg:energy}
$x^e[0] \gets x[0]$\;
\For{$t \in 0, \ldots,$ number of steps $-1$}{
    Generate $x[t+1]$ from $x[t]$ using one step of our model in Eq.~\ref{eq:ds} with no inputs\;
    $E[t] \gets \|x[t+1] - x[t]\|_2^2$\;
    $x^e[t+1] \gets x[t] - \alpha \frac{\partial E(t)}{\partial x[t]}\Big|_{x[t]}$\;
    $x[t+1] \gets x^e[t+1]$\;
}
\end{algorithm}

Here, $\alpha= 0.1$ and the number of steps is 1000. $[x^e[0], x^e[1], ..., x^e[1000]]$ gives a single energy minimization trajectory. Final states $x^e[1000]$ for each initialization were visualized as red crosses using PCA.

\subsection{Visualization analysis}
To visualize attractor manifolds, we reduced the dimensionality of $x[t]$ and $x^e[t]$ into 3 with Principle Component Analysis (PCA). To fit the data to PCA, we first fed 100 trials from the test set to the \textit{fully-trained PLRNN} and obtained the corresponding activations $x[t]$s. All the states in $x[t]$s are used to fit the PCA. The resultant PCA model transformed all the $x^e[t]$s and $x[t]$s obtained from checkpoint models during the training.

\subsection{Latent circuit analysis in rank-one recurrent neural networks} \label{app:rank-one}

We investigated rapid skill acquisition mechanisms using interpretable rank-one recurrent neural networks (RNNs). The network dynamics are governed by:

\begin{equation}
    \tau \dot{r}(t) = -r(t) + \tanh(W r(t) + C u(t) + h)
\end{equation}
where $r(t) \in \mathbb{R}^N$ represents neural activity, $W \in \mathbb{R}^{N \times N}$ is the recurrent weight matrix, $C \in \mathbb{R}^{N \times 2}$ is the input weight matrix, $u(t) \in \mathbb{R}^2$ is the input, and $h \in \mathbb{R}^N$ is a bias term. We constrained $W$ to be rank-one: $W = \frac{1}{N}mn^T$, with $m, n \in \mathbb{R}^N$. This allows definition of a one-dimensional latent variable $\kappa(t) = \frac{n^T r(t)}{N}$, following:
\begin{equation}
    \tau \dot{\kappa}(t) = -\kappa(t) + \frac{n^T}{N}\tanh(m\kappa(t) + C u(t) + h)
\end{equation}
The network had 2 input units (random signal and pulse signal) and 1 output unit. We trained 100 rank-one RNNs ($N = 40$ neurons) for 5000 epochs each. Training parameters were as follows:
\begin{itemize}
    \item Optimizer: Adam with initial learning rate 0.05
\item Learning rate scheduler: rate halved every 500 epochs without validation accuracy improvement
\item Batch size: 1000
\item Loss function: Mean squared error
\item Time constant $\tau = 10$ms
\end{itemize}
We analyzed latent circuit dynamics across training epochs by computing:
\begin{itemize}
    \item Test accuracy: Fraction of correct trials (predicted sum within 0.04 of true sum) on a held-out test set of 5000 trials.
\item Number of fixed points: We calculated $\dot{\kappa}$ for $\kappa \in [-3, 3]$ (100 equally spaced points) and counted zero-crossings.
\item Flatness score: We quantified approximate line attractors by measuring the total length of $\kappa$ intervals where both $|\dot{\kappa}| < 0.2$ and $|\partial_\kappa[\dot{\kappa}]| < 0.2$.
\end{itemize}
Metrics were aligned across networks based on the end of the search phase.

\subsection{Stimulus Decoding Experiments} \label{app:decoding}

To test the memory capabilities of attractor-incentivized networks compared to unregularized PLRNNs, we conducted cue and stimulus decoding experiments. We picked 10 unregularized PLRNNs and 10 PLRNNs trained with MAR and MAI for the delayed addition task from the networks shown in Fig. \ref{fig:figs4}A: $T= 60$.

We first created a new addition dataset with predetermined cue positions (i.e. $t_1 = 5$ and $t_2 = \frac{T}{2} - 1$). The goal was to quantify networks ability to predict $u^1_{t_1}, u^1_{t_2}$, and $u^1_{t_1} + u^1_{t_2}$ from instantaneous activations of all the neurons ($x_{S}[t] = {x_i[t]}_{i= \{1, ..., N\}}$), memory neurons ($x_{S_1}[t] = {x_i[t]}_{i= \{1, ..., N_{reg}\}}$), and computational neurons ($x_{S_2}[t] = {x_i[t]}_{i= \{N_{reg}+1, ..., N\}}$) separately. Analysis was conducted for all the time steps, $t \in {1, ..., T}$. 

To predict cues and/or targets from neuronal activity, we used linear regression models. For a given network and a time step $t$, 9 linear regression models were independently fitted to map 3 types of instantaneous inputs (i.e. $x_S[t], x_{S_1}[t], x_{S_2}[t]$) into 3 types of decoding targets (i.e. $u^1_{t_1}, u^1_{t_2}$, and $u^1_{t_1} + u^1_{t_2}$). We repeated the procedure for $t \in \{1, ..., T\}$ and all 20 networks (\textit{i.e.}, 10 unregularized PLRNNs and 10 regularized PLRNNs). All the fitting was done on the training dataset with $10,000$ trials and testing was done on the testing dataset with $1,000$ trials.

\subsection{Frequency Analysis}

We analyzed frequency distributions of $x[t]$ belonging to unregularized PLRNNs, PLRNNs with MAI, and PLRNNs with MAI+MAR. To conduct this analysis, we selected 3 representative networks from Fig. \ref{fig:figs4}\textbf{A}, with $T= 20$. We first concatenated 50 trials with $T= 20$ from the test set and obtained a long trial with a temporal length of $T= 1000$. Activations $x[t]$s were obtained for this long input trial. Obtained activations for 10 neurons are shown in Fig. \ref{fig:figs4}\textbf{C}. 

To obtain the frequency plots, we separately considered activities, $x[t]$, of computational and memory neurons. We first divided the trajectories of the neuronal activations by the maximum activation of each trajectory. Fast Fourier transform followed by frequency shift was then applied. We then obtained absolute values from the resultant frequency sequence. The resultant frequency sequences were divided again by the maximum frequency component of the corresponding frequency sequence. Finally, we took the mean frequency sequence over neurons. Resulting frequency plots are shown in Fig. \ref{fig:figs4}\textbf{C}.

\subsection{Training speed during the search phase}

We calculated the training speed during the search phase by taking $\frac{1}{\text{epoch}_b}$. $\text{epoch}_b$ is computed based on the stalling points in the loss function during training. Specifically, we computed the first epoch that achieves a particular value in the loss or accuracy levels, after which rapid learning took place. For the delayed addition task, this was $\sim 0.08$ for the fraction of correct trials. For the K-bit flip flop task, this was $\sim 0.5$ in the correlation values between the target and the outputs. For the evidence accumulation task, this was $\sim 0.3$ in the mean squared distance between target and reconstructed coherence. We performed visual inspection in all cases to ensure that we captured the onset of the rapid comprehension phase.

\subsection{Cue-responsive fixed point generation (online and local)}
\label{sec:onlinelearning}
For our online experiments, we once again used leaky firing rate RNNs:  
\begin{equation} \label{eq:generator}
\begin{split}
\tau \frac{dr_i(t)}{dt} &= -r_i(t) + f(z_i(t)), \\
z_i(t) &= \sum_{j=1}^{N_{\rm rec}} W_{ij} r_j(t) + \sum_{j=1}^{N_{\rm in}} C_{ij} s_j(t) + \epsilon_i(t),\: \forall i \in [N_{rec}].
\end{split}
\end{equation}
Here, $r_i$ is the activity or firing rate of neuron $i$ and $z_i$ is the total input current to neuron $i$.  $N_{rec}$, the number of neurons, is 400 and $N_{in}$, the number of input channels. The rest is defined as before. Furthermore, the time constant $\tau = 10 ms$, and $f(\cdot) = \tanh(\cdot)$. Each element of $W$ was initially drawn from a normal distribution with mean 0 and standard deviation $1.8/\sqrt{N_{rec}}$ while each element of $C$ was drawn from a standard normal. To prevent self-excitation, we also enforced $W_{ii} = 0,\: \forall i \in [N_{rec}]$. 

While the continuous time formulation provides a theoretically motivated window into the time dynamics of the network, the simulations were performed via discretization with $\Delta t$ time steps. The discretized network dynamics followed:
\begin{equation}
 r_{i}[t + \Delta t] = (1-\alpha)  r_{i}[t] + \alpha f( z_{i}[t + \Delta t]),
\end{equation}
where $\alpha = \frac{\Delta t}{\tau}$ is the unitless normalized discretization time, here chosen to be $\alpha = 0.1$.

The learning process began after $s[t]$ becomes nonzero, \textit{i.e.}, the arrival of the rectangular pulse.  After completion of this pulse, we enforced that the network aims to produce its previous activity $r[t - \Delta t]$. Namely, we performed a local gradient step on the loss function:
\begin{equation}
\lambda(r_i[t] - r_i[t-\Delta t])^2
\end{equation}
for all neurons, where $\lambda$ determines the strength of this derivative regularizer. A step of this update rule, after lumping the derivative of the non-linearity into the regularization strength $\lambda \to \lambda_{\rm TCR}$, corresponded to
\begin{equation}
W[t+\Delta t] =W[t] - \lambda_{\rm TCR} \left(r[t]-r[t-\Delta t]\right)r[t-\Delta t]^T,
\end{equation}
where set $\lambda_{\rm TCR} = 1/1000$. We repeated this update several times, in a randomized manner across cues, as shown in Fig. \ref{fig:fig7}. Each pass across all cues was considered a single epoch. For Fig. \ref{fig:fig7}\textbf{C}, we computed the Jacobian $J$ at some given $r_*$ by taking partial derivatives of the right-hand side in Eq. (\ref{eq:generator}).

\section{Supplementary Figures}
For the supplementary figures referenced in the main text, please see below.

\clearpage
\setcounter{figure}{0}
\renewcommand{\thefigure}{S\arabic{figure}}
\renewcommand{\theHfigure}{S\arabic{figure}}
\setcounter{table}{0}
\renewcommand{\thetable}{S\arabic{table}}

\begin{figure*}
    \centering
    \includegraphics[width=\textwidth]{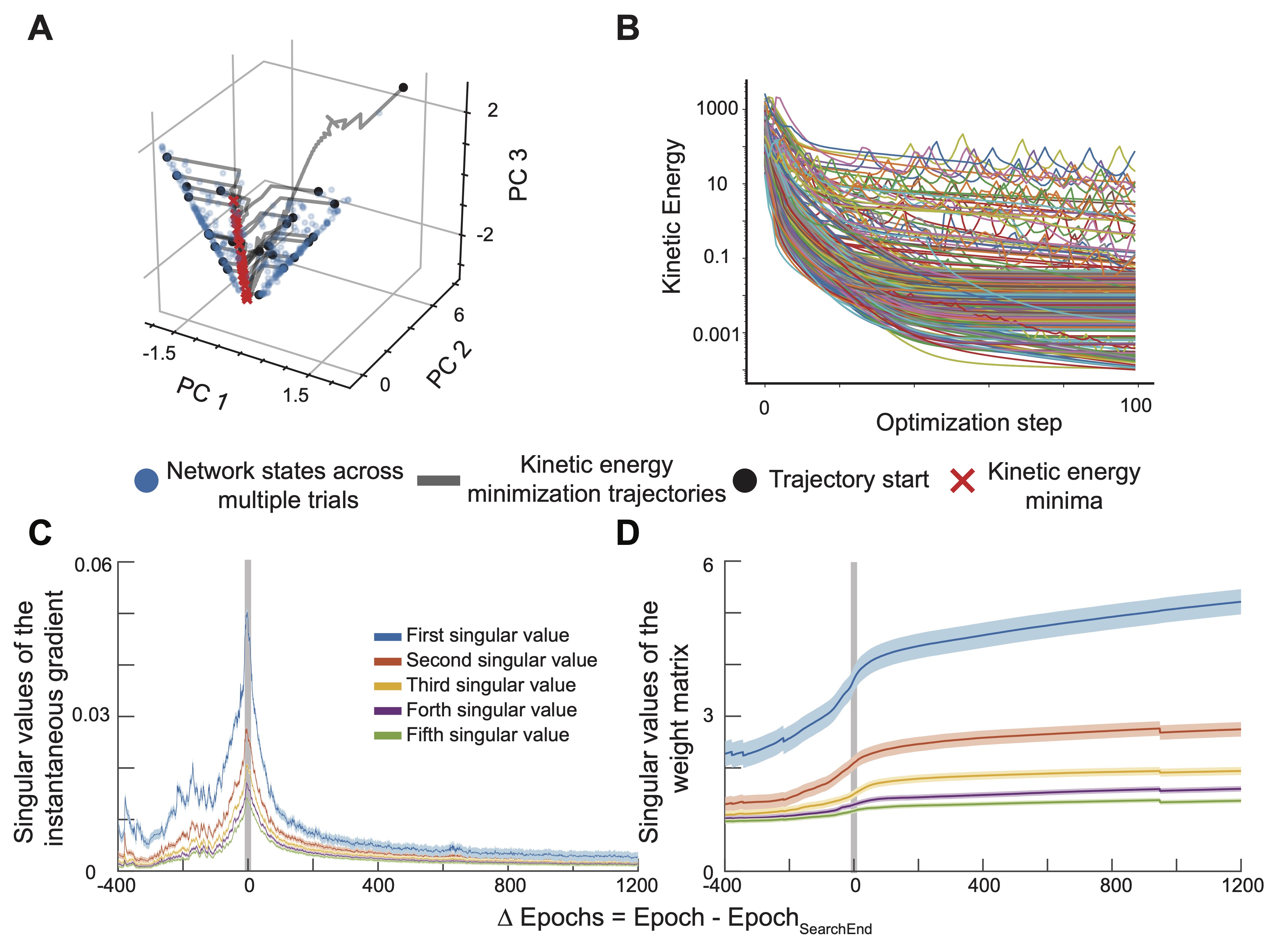}
    \caption{\textbf{Abrupt changes in the slow point landscapes are accompanied with the destabilization of the gradients.} \textbf{A-B} We studied the slow point landscapes of the networks during training, which are computed using the energy minimization process. \textbf{A} Since the kinetic energy minimization is a non-convex problem, the optimization trajectories (black lines) and the converged slow points (red crosses) depend on the initial points, which are chosen randomly from the neural states visited during the trials (blue dots). \textbf{B} The kinetic energy is plotted over optimization step, each line corresponds to a different trajectory from a distinct initialization. Though the kinetic energy decreases over iterations, it does not reach zero (or close to the machine precision), indicating the existence of slow, but not necessarily fixed, points. \textbf{C-D} The gradients first destabilize then recover during the auto-catalytic changes, but the weight changes are permanent. \textbf{C} During the geometric restructuring in RNNs' phase spaces, the weights undergo large updates. \textbf{D} The weight matrices evolve abruptly during GR events and settle into new equilibria after completion. Solid lines: means. Shaded areas: s.e.m. over 19 networks ($N= 40$ neurons, $T= 40$ time steps, $3000$ epochs training, no regularization).}
    \label{fig:figs1}
\end{figure*}

\begin{figure*}
    \centering
    \includegraphics[width=\textwidth]{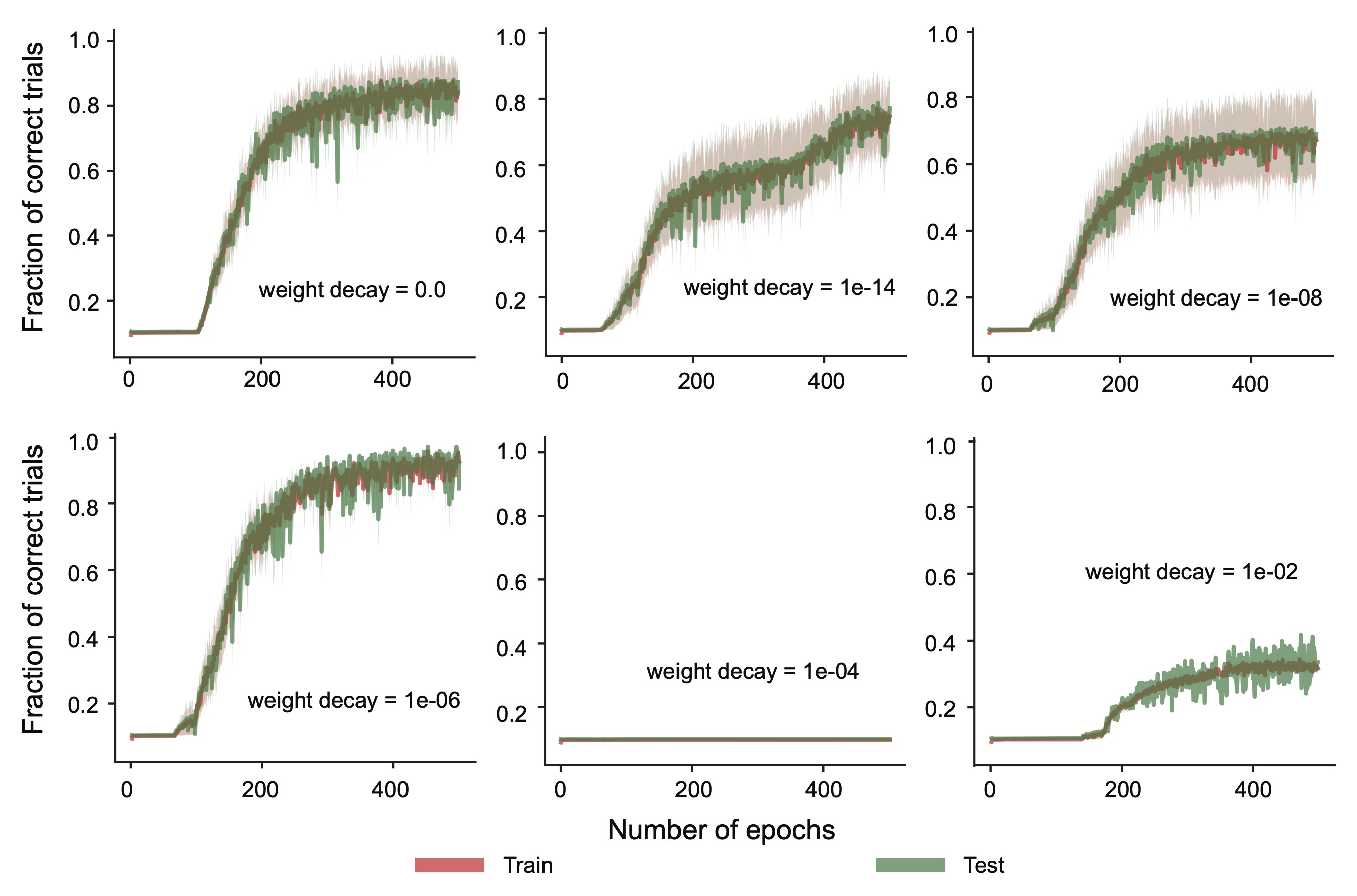}
    \caption{\textbf{Weight decay does not have a consistent effect on the duration of the search phase.} To test whether weight decay can facilitate the desired geometric restructuring, we trained PLRNNs ($N= 40$) regularized with varying levels of weight decay values on delayed addition tasks with $T=40$. We had not observed any consistent affect of the weight decay values on the duration of the search phase. Solid lines: mean. Shaded areas: s.e.m. over 10 networks.}
    \label{fig:figs2}
\end{figure*}

\begin{figure*}
    \centering
    \includegraphics[width=\textwidth]{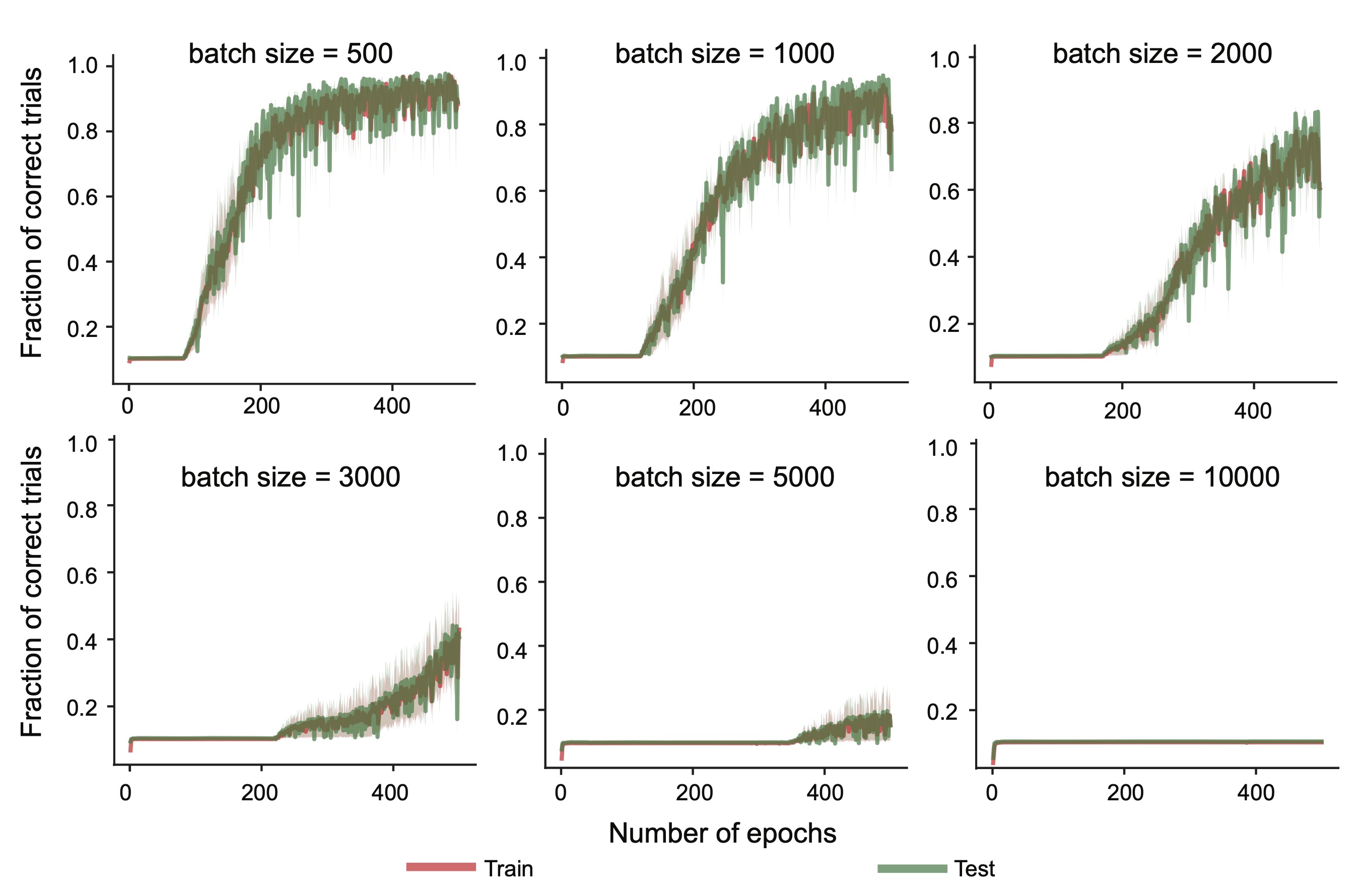}
    \caption{\textbf{Increasing the stochasticity of the gradient shortens the search phase.} Similar to Fig. \ref{fig:figs2}, we tested the effects of stochasticity in training by varying the batch sizes while training unregularized PLRNNs ($N= 40$) to perform delayed addition tasks with $T=40$. Increased stochasticity had a beneficial effect on the training by speeding up the necessary geometric restructuring. Solid lines: means. Shaded areas: s.e.m. over 5 networks.}    
    \label{fig:figs3}
\end{figure*}

\begin{figure*}
    \centering
    \includegraphics[width=\textwidth]{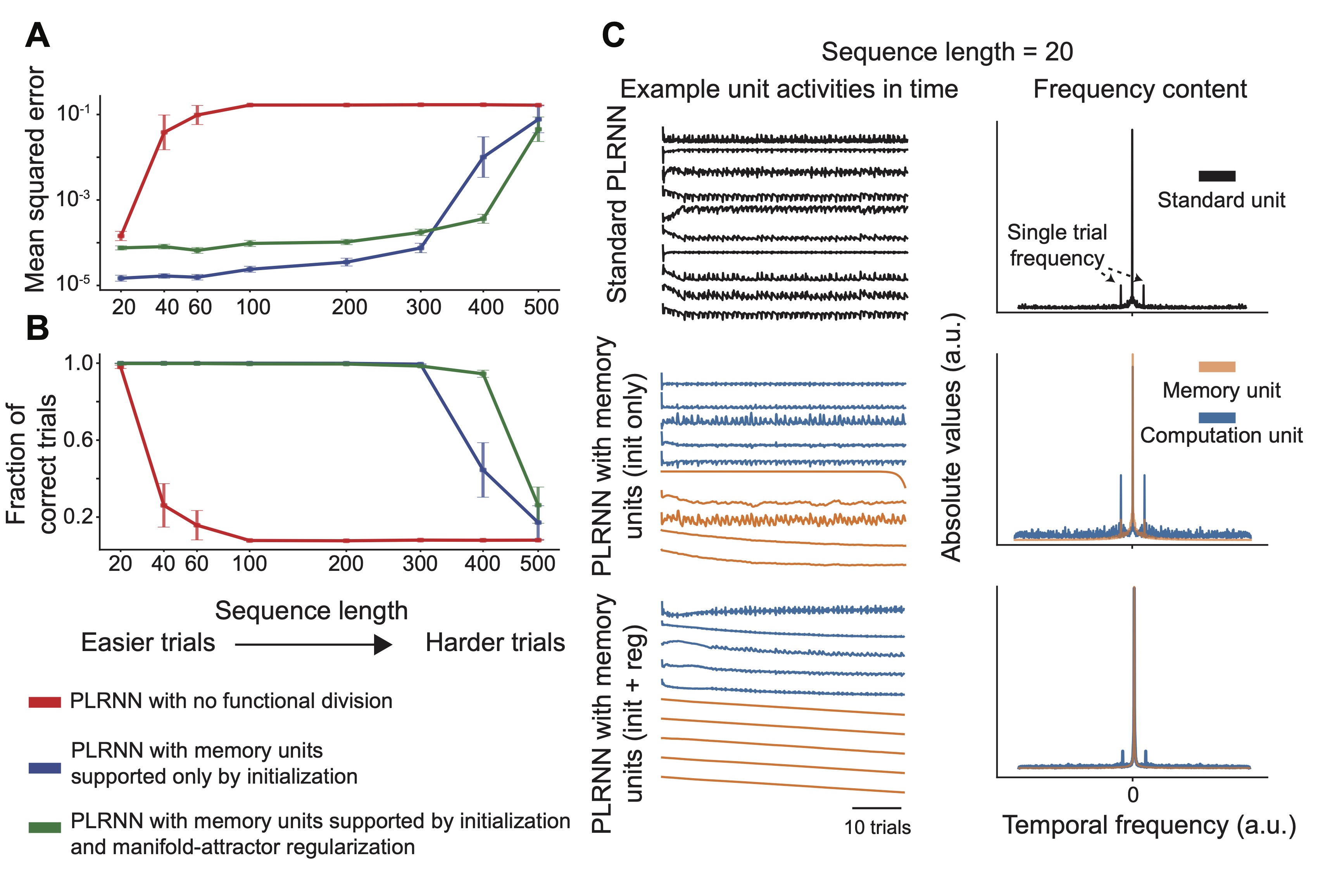}    \caption{\textbf{Manifold attractor initialization and regularization incentivize slow time dynamics}. \textbf{A-B} We trained unregularized and regularized PLRNNs on the delayed-addition task with varying levels of sequence length. We plotted (\textbf{A}) mean squared errors and (\textbf{B}) the fractions of correct trials. Solid lines: means. Error bars: s.e.m. over 10 runs. \textbf{C} Analysis of frequency content in PLRNNs' computation and memory units. Memory units in networks with MAR/MAI show slow time dynamics, whereas the neural activities of the computation units have high frequency components. For all experiments, MAR/MAI was applied to $20$ out of $40$ neurons. MAR regularization weight was $\lambda_{\rm MAR}=5$.} 
    \label{fig:figs4}
\end{figure*}

\begin{figure*}
    \centering
    \includegraphics[width=\textwidth]{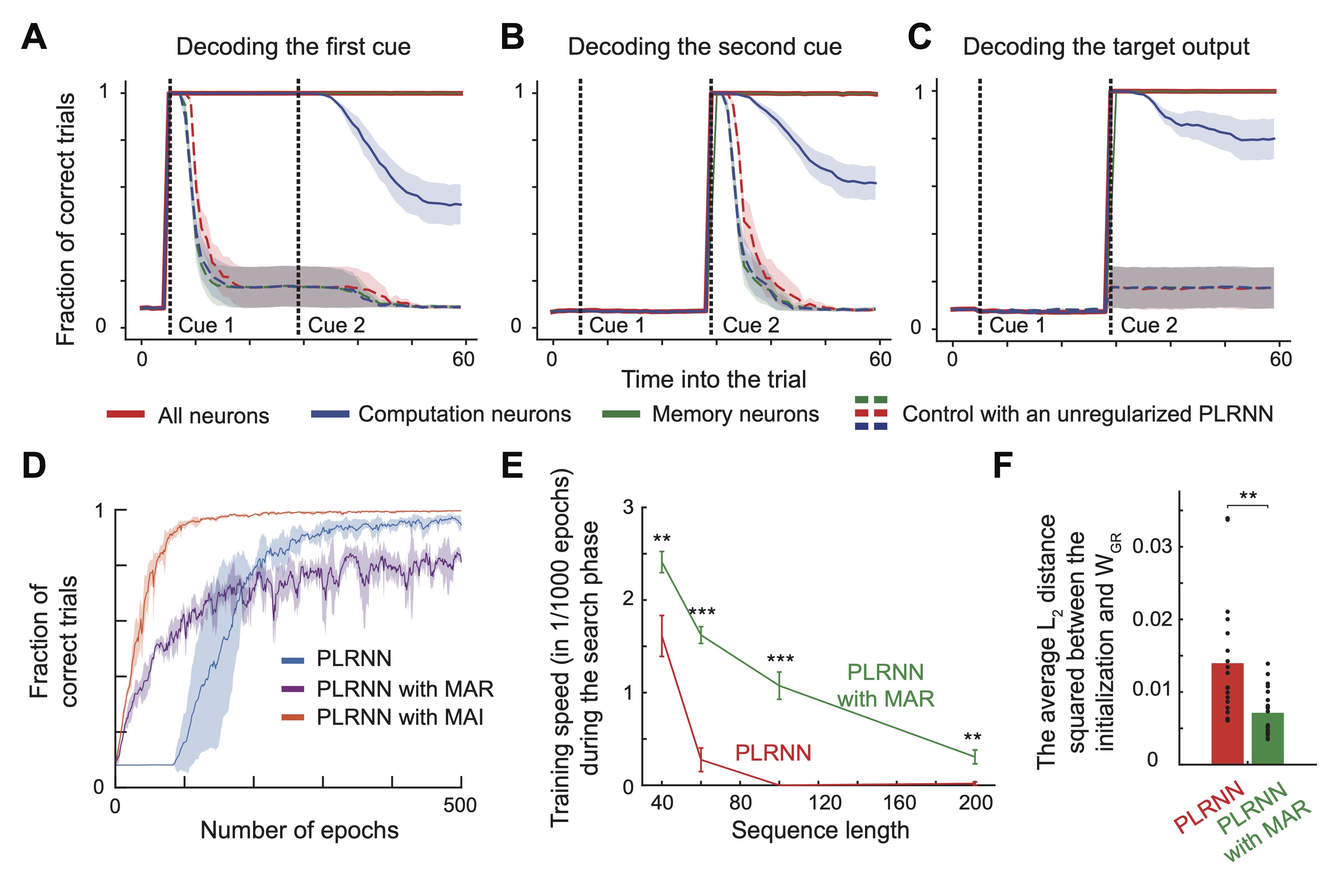}
    \caption{\textbf{Manifold attractor initialization and/or regularization enhance the memory capabilities of PLRNNs by choosing a closer $W_{\rm GR}$.}  \textbf{A-C} To test the memorization of the task-relevant information in the unit activations, we trained PLRNNs on delayed addition tasks with $T=60$, which were initialized with manifold attractors and had incentives to retain them due to a regularization term (Appendix \ref{app:mar}). Here, $20$ out of $N=40$ neurons were regularized with $\lambda_{\rm MAR}=5$. During testing only, we focused on a variation of the delayed addition task, in which the input times for the cues were fixed (See Appendix \ref{app:decoding}). Using the activities of memory, computation, and all neurons, we trained linear estimators for the first cue \textbf{(A)} , the second cue \textbf{(B)}, and the target output \textbf{(C)}, \textit{i.e.}, the sum of the two cues. Computation, but not the memory, neurons had reduced memory of cue 1, cue 2, and the output. As a control (dashed lines), we also trained unregularized PLRNNs, which were not able to retain the memories for either of the cues to begin with. Thus, the emerging manifold attractors in the memory neurons enhance the short-term memory capabilities. Solid and dashed lines: means. Shaded areas: s.e.m. over $10$ networks. \textbf{D} MAI creates a plane attractor using a small subset of units in the PLRNN (Appendix \ref{app:mar}), which gives the network enhanced memory capabilities. Therefore, MAI practically skips the search phase while having an extremely abrupt learning phase. In comparison, MAR encourages the plane attractor formation during the learning with a regularization term, leading to undetectable fast search phase. Solid lines: means. Shaded areas: s.d. over 5 networks that are trained on the delayed addition task with $T=40$ and $N=40$, with $20$ regularized memory neurons whenever applicable. MAR regularization weight was $\lambda_{\rm MAR}=5$. \textbf{E} MAR leads to a faster training speed during the search phase. The comparisons are performed with two-sided Wilcoxon signed-rank tests ($^{***}p<10^{-3}$, $^{**}p<10^{-2}$). Solid lines: means. Error bars: s.e.m. over 20 networks. For MAR, we used $\lambda_{\rm MAR} = 0.1$ and regularized half of the $N=40$ neurons. \textbf{F} PLRNNs trained with MAR show lower $L_2$ distances between initialization and $W_{\rm GR}$, indicating that regularized networks cross the desired weight subspace boundary at a closer parameter configuration $W_{\rm GR}$. The comparison is performed with a two-sided rank sum test ($^{**}p<10^{-2}$).} 
    \label{fig:figs5}
\end{figure*}

\begin{figure*}
    \centering
    \includegraphics[width=\textwidth]{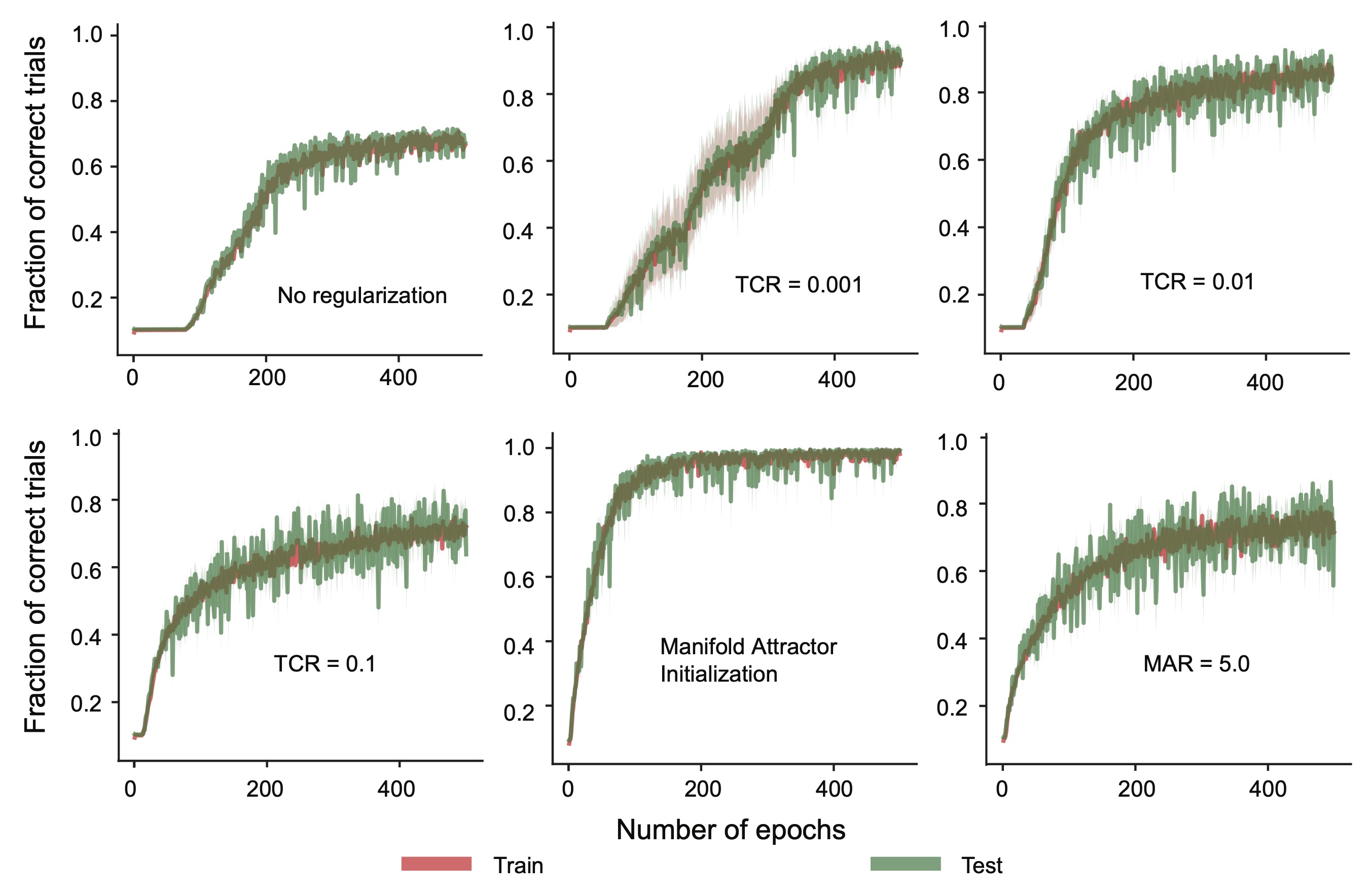}
    \caption{\textbf{Enforcing temporal consistency on subset of neurons shortens the search phase.} To test the efficacy and efficiency of TCR, we trained PLRNNs (regularized $N_{\rm reg}=20$ out of $N=40$ neurons) to perform delayed addition tasks with $T = 40$. Higher TCR strength led to faster conclusion of the search phase, whereas PLRNNs trained with either MAR or MAI practically skipped the search phase. Solid lines: means. Shaded areas: s.e.m. over 10 networks.}
    \label{fig:figs6}
\end{figure*}

\begin{figure*}
    \centering
    \includegraphics[width=0.9\textwidth]{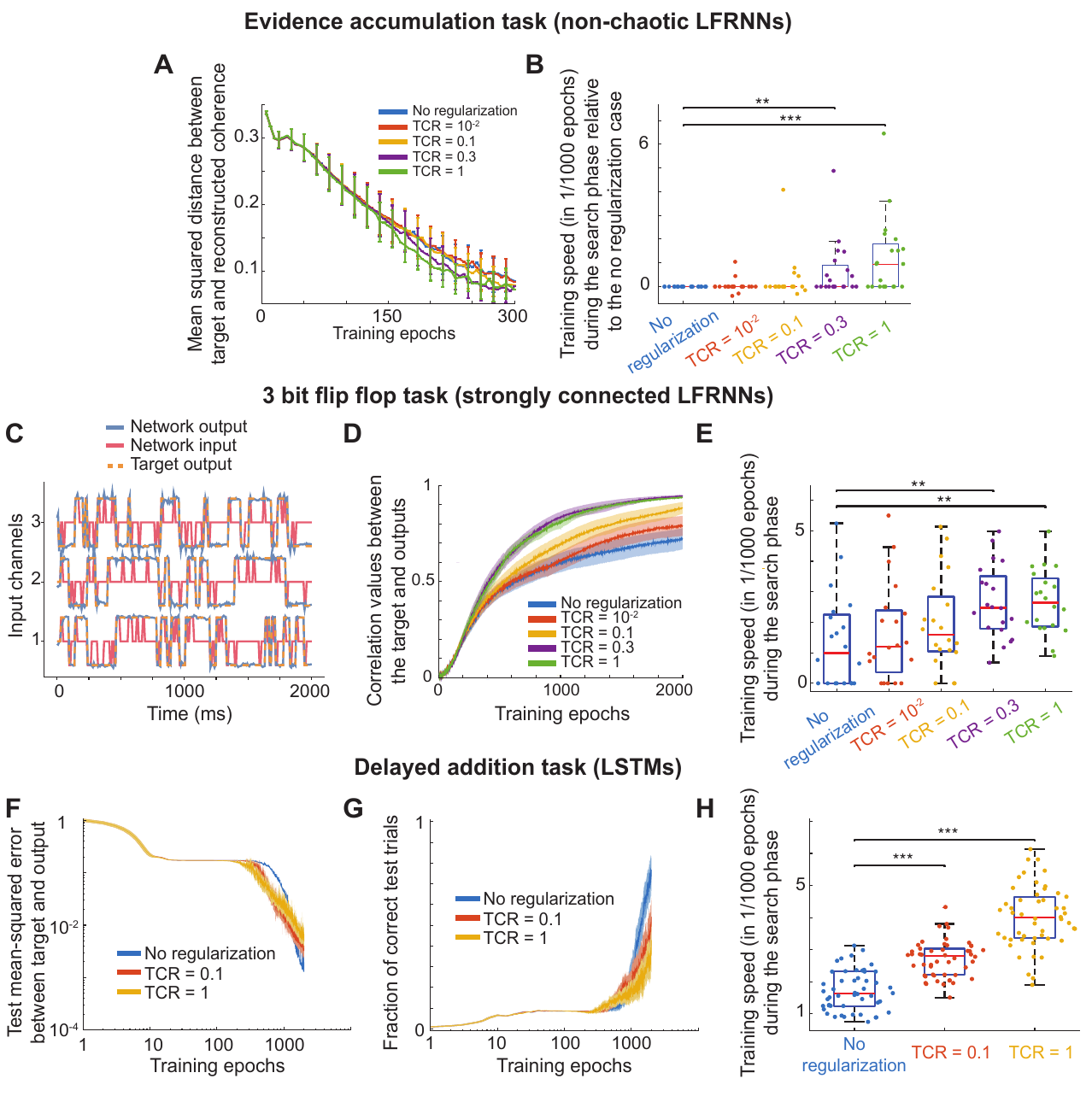}
    \caption{\textbf{TCR allows abrupt training in diverse network architectures and short-term memory tasks.} \textbf{A-B} To show that TCR improves the convergence of non-chaotic RNNs, we trained the networks on the same task as in Fig. \ref{fig:fig6} but with Xavier initialization and for 300 epochs. Similar to before, TCR resulted in slightly lower loss \textbf{(A)} and faster training speed during the search phase \textbf{(B)}. The lfRNNs had $N=50$ neurons, $N_{\rm reg} = 25$ were regularized with TCR whenever applicable, the neuronal decay time was $\tau = 10ms$, and the lfRNNs were discretized with $\Delta  = 8ms$. The comparisons were performed with two-sided Wilcoxon signed-rank tests ($^{**}p<10^{-2}$ and $^{***}p<10^{-3}$).  \textbf{C-E} To test whether time consistency regularization improves training of fixed points in strongly connected networks, we trained 20 lfRNNs ($N=50$, $N_{\rm reg} = 25$, $\tau = 10ms$, $\Delta t = 8ms$, and $g=3$) on a 3-bit flip-flop task for $2,000$ epochs. These networks, in the limit $N \to \infty$ are known to show chaotic behavior \cite{sompolinsky1988chaos}. \textbf{C} In the flip-flop task, the network receives three (mostly zero) inputs, with occasional positive and negative pulses signalling state changes. The network should adapt its three outputs to the latest state presented by the input, which requires switching its internal dynamics between $2^3 = 8$ distinct states. As before, TCR enhanced  the learning speed and performance (\textbf{D}) and accelerated training during the search phase (\textbf{E}). \textbf{F-H} To test whether time consistency regularization improves training speed of LSTMs during the search phase, we trained 50 LSTMs on the delayed addition task with $T=40$. Here, $N_{\rm reg}= 20$ out of $N=40$ neurons were regularized with TCR. The error values (\textbf{F}) and the accuracies (\textbf{G}) over $1,000$ test trials as a function of number of training epochs. Solid lines: means. Shaded regions: 95\% confidence intervals across $50$ networks. \textbf{H} Consistent with all our results, TCR led to faster search phase with LSTMs as well. The comparisons were performed with two-sided Wilcoxon signed-rank tests ($^{***}p<10^{-3}$).} 
    \label{fig:figs7}
\end{figure*}

\end{document}